# Wiener Reconstruction of Density, Velocity, and Potential Fields from All-Sky Galaxy Redshift Surveys

K. B. Fisher[1], O. Lahav[1], Y. Hoffman[2], D. Lynden-Bell[1], & S. Zaroubi[2]
[1] *Institute of Astronomy, Madingley Rd., Cambridge, CB3 0HA, England*
[2] *Racah Institute of Physics, The Hebrew University, Jerusalem, Israel*

3 June 1994

**ABSTRACT**
We present a new method for recovering the cosmological density, velocity, and potential fields from all-sky redshift catalogues. The method is based on an expansion of the fields in orthogonal radial (Bessel) and angular (spherical harmonic) functions. In this coordinate system, peculiar velocities introduce a coupling of the radial harmonics describing the density field in redshift space but leave the angular modes unaffected. In the harmonic transform space, this radial coupling is described by a distortion matrix which can be computed analytically within the context of linear theory; the redshift space harmonics can then be converted to their real space values by inversion of this matrix. Statistical or "shot" noise is mitigated by regularizing the matrix inversion with a Wiener filter (roughly speaking, the ratio of the variance in the signal to the variance of the signal plus noise). The method is non-iterative and yields a minimum variance estimate of the density field in real space. In this coordinate system, the minimum variance harmonics of the peculiar velocity and potential fields are related to those of the density field by simple linear transformations. Tests of the method with simulations of a Cold Dark Matter universe and comparison with previously proposed methods demonstrate it to be a very promising new reconstruction method for the local density and velocity field. A first application to the 1.2 Jy *IRAS* redshift survey is presented.

## 1 INTRODUCTION

Historically, redshift surveys have provided the raw data and impetus for much of the research into the nature of the three dimensional galaxy distribution. Redshift surveys provide invaluable qualitative cosmographical information and have revealed a wide variety of structure including voids, filaments, and superclusters. Redshift surveys have also greatly increased our quantitative understanding of the galaxy distribution as is evident in the wealth of statistical measures that have been applied to them over the last two decades.

Today, cosmologists have at their disposal not only large, near full sky, maps of the galaxy redshift distribution but also a growing database of direct measurements of galaxy distances (e.g., Burstein 1990; Bertschinger 1992, and references therein). A large and complete sample of galaxy distances would represent a marked improvement over redshift surveys for measuring the properties of the galaxy distribution since the clustering pattern in redshift space can be systematically different from the true clustering in real space (Kaiser 1987). Unfortunately, despite the heroic efforts of researchers in the field, accurate distances are still lacking for the vast majority of galaxies with redshifts. Fortunately, however, it is possible to *reconstruct* the peculiar velocity field, and hence distance estimates, using only redshift data if one is willing to make assumptions about the underlying dynamics which have given rise to the velocities.

In the linear theory of gravitational instability, the gravity and velocity vectors are parallel and related to each other by a proportionality constant which depends only on the mean mass density of the universe, $\Omega$. If one makes the further (and crucial) assumption that the galaxy distribution traces the underlying mass distribution, then it is possible to determine the gravity field from the galaxy distribution and thus derive the linear velocity field. Agreement between the predicted and directly measured peculiar velocities would give strong evidence for the reality of distance indicators and would allow a measurement of $\Omega$. Indeed measurements of this kind have been made using large redshift surveys obtained from both *IRAS* and optical databases (e.g. Strauss & Davis 1988; Kaiser *et al.* 1991; Yahil *et al.* 1991, hereafter YSDH; Hudson 1994; Freudling *et al.* 1994).

There are several complications, however, in reconstructing the peculiar velocity field directly from redshift surveys. First, as already mentioned, one must adopt some prescription for relating the fluctuations in the galaxy distribution to those in the mass. The standard approach is to assume that the two are related by a proportionality constant, the so-called linear bias parameter ($b$), which is independent of scale (this model is motivated more by simplicity than by physical principles). Even if galaxies faithfully trace the mass, there remains a dynamical mapping problem which arises if one attempts to equate redshifts with the true distances since the gravity field derived from the distribution of galaxies in redshift space will be *systematically* different from the true density field. In the absence of statistical noise and nonlinear evolution, this mapping is well-defined and described by the linear theory (Kaiser 1987). Moreover, linear theory can be applied directly in redshift space by appropriately modifying the Poisson equation to self-consistently account for the effect of redshift distortion (Nusser & Davis 1994; hereafter ND). To date, the majority of reconstruction techniques have been based on linear theory with the added assumption of linear biasing (see Dekel 1994 and references therein).

In practice, however, there are several problems that are encountered when linear theory is used for the reconstruction. First, the simple proportionality between the gravity and velocity fields in linear theory is only valid when the density fluctuations are small. Once the clustering has gone nonlinear this one-to-one correspondence will be erased due to shell crossing. Another problem arises from the finite sampling of existing redshift catalogues both in the form of incomplete sky coverage and limited depth. The problem of sky coverage has been improved substantially through the use of infrared selected surveys which are less affected by Galactic obscuration, however even these samples leave $\gtrsim 10\%$ of the sky unobserved. In addition, redshift catalogues based on the *IRAS* database are flux-limited and consequently the number density of galaxies declines sharply with distance. Inevitably, the estimate of the density field becomes subject to large statistical uncertainties at large distances. The necessary procedure needed to remedy the problems caused by nonlinear evolution and incomplete sampling is to smooth the density field. Consequently, it is desirable to implement a smoothing algorithm which is motivated by both the clustering properties of the galaxy distribution and the sampling of the catalogue.

The reconstruction method presented in this paper is based on linear theory. The main innovation in our method is the way in which smoothing is performed. Our algorithm is based on a "Bayesian" method where a *prior* model for the galaxy distribution is assumed. The advantage of assuming a prior is that it allows the significance of the statistical noise to be evaluated and then "cleaned" from the reconstruction. This filtering is accomplished by a Wiener filter (Wiener 1949; hereafter WF) which is roughly the ratio of the variance in the signal to the variance in the signal plus noise. A related Bayesian reconstruction method was applied by Kaiser & Stebbins (1991; see also Stebbins 1994) to derive maps of the peculiar velocity field.

The WF is optimal in the sense that the variance between the derived reconstruction and the underlying true field is minimized. WF reconstructions provide a noted improvement over previous methods in that the degree of smoothing has a physical justification, as determined from the prior; the reconstruction works directly on the raw galaxy data rather than a smoothed distribution. In the limit of high signal to noise ratio, the WF acts as a straight inversion; however, the WF acts to suppress the contribution of data strongly contaminated by noise.

The WF approach has recently been applied to several reconstruction problems in large scale structure (LSS). Lahav *et al.* (1994; hereafter LFHSZ) reconstructed the angular distribution of *IRAS* galaxies while Bunn *et al.* (1994) applied the method to the temperature fluctuations in the *COBE* DMR maps. A nice review of the WF and linear estimation can be found in Rybicki & Press (1992). A detailed overview of the concept and theory of Wiener filtering as it pertains to LSS reconstructions can be found in Zaroubi *et al.* (1994; hereafter ZHFL).

In this paper, we outline a method for reconstructing of the density, velocity, and potential fields by invoking the WF in the transform space of spherical harmonics. A similar method based on a Cartesian representation is employed by Ganon & Hoffman (1993) and Bistolas, Zaroubi, & Hoffman (1994, in preparation, see also Hoffman 1994). Analysis of redshift surveys in spherical coordinates turns out to be very convenient for addressing many of the problems mentioned above. Incomplete sky coverage can be described in terms of an angular "mask" while the sampling in a flux-limited survey is characterized by a radial selection function; moreover distortions induced by the peculiar velocity field affect only the radial coordinates of the galaxies.

Spherical Harmonic Analysis (SHA) has been discussed for analysing projected surveys about 20 years ago (Peebles 1973), but was not very useful given the poor sky coverage of the samples existing at the time. More recently SHA has been reconsidered and applied to *IRAS* selected galaxy surveys (Fabbri & Natale 1989; Scharf *et al.* 1992; Scharf & Lahav 1993, Fisher, Scharf, & Lahav 1993 (hereafter FSL), Scharf 1993; Lahav 1994) and to the peculiar velocity field (Regös & Szalay 1989, Lynden-Bell 1991, Lahav 1992, Fisher 1994). The present paper extends the theoretical formalism of ZHFL to model self-consistently the effects of redshift distortion and to give a unified method for reconstructing cosmological fields.

We begin in § 2 with a brief overview of how the density field can be decomposed into a set of orthogonal spherical harmonics and show how these harmonics can be estimated from redshift catalogues. In § 3, we show how these harmonics are related to the fluctuations of the galaxy distribution in *real* space. In § 4, we describe a method for reconstructing the real space harmonics and show how these can be used to recover the peculiar velocity and potential fields. This method involves a dynamical mapping which is supplemented with the Wiener filtering of the harmonics to minimize statistical noise in an optimal way. We describe a practical implementation of our reconstruction technique in § 5 and test it with a series of mock galaxy catalogues extracted from numerical simulations of a Cold Dark Matter (CDM) universe. We compare our reconstructions of the peculiar velocity field with several other methods in the literature in § 7.2. We conclude in § 7.3. The majority of the mathematical derivations have been relegated to a set of Appendices in an effort to make the main text more readable.

## 2 EXPANSION OF DENSITY FIELD IN SPHERICAL COORDINATES

We begin, therefore, by expanding the density field, $\rho(\mathbf{r})$, within a spherical volume of radius $R$ in a Fourier-Bessel series (Binney & Quinn 1991; Lahav 1994):

$$\rho(\mathbf{r}) = \sum_{lmn} C_{ln}\, \rho_{lmn}\, j_l(k_n r)\, Y_{lm}(\hat{\mathbf{r}}) \quad , \tag{1}$$

where here and throughout the summation sign denotes the following:

$$\sum_{lmn} = \sum_{l=0}^{l_{max}} \sum_{m=-l}^{+l} \sum_{n=1}^{n_{max}(l)} \quad . \tag{2}$$

This expansion is the spherical analog to the well-known Fourier decomposition into plane waves which is conventionally employed in analyses of large scale structure. In Equation 1, $j_l(x)$ is a spherical Bessel function, $Y_{lm}(\hat{\mathbf{r}})$ is the spherical harmonic corresponding to the angular coordinates, $\hat{\mathbf{r}}$, and $C_{ln}$ is a normalization constant to be defined below. The Fourier-Bessel expansion forms a complete set in the limit that $l_{max}$ and $n_{max}(l)$ tend to infinity. Obviously, in practice the summations are truncated at a finite number of angular and radial modes. This truncation will limit the resolution of the density field on small scales; the choice of $l_{max}$ and $n_{max}(l)$ is addressed in Appendix B.

The discrete spectrum of radial modes, denoted by the index $k_n$, arises because we impose a boundary condition at the edge of the spherical volume at $r = R$. In this paper we have demanded that the logarithmic derivative of the gravitational potential be continuous at the boundary; this condition is satisfied for all $l$ if $j_{l-1}(k_n R) = 0$ ($l = 0$ poses no problem since the recursion relations define the Bessel functions of negative order, see Abramowitz & Stegun, §10.1.12); a short derivation of this result, as well as a discussion of other possible boundary conditions, is given in Appendix A.

With our choice of boundary conditions, the Bessel and spherical harmonic functions form an orthogonal set of basis functions. By orthogonality, we mean that

$$\int_{4\pi} d\Omega\, Y_{lm}(\hat{\mathbf{r}})\, Y^*_{l'm'}(\hat{\mathbf{r}}) = \delta^K_{ll'} \delta^K_{mm'}$$

$$\int_0^R dr\, r^2\, j_l(k_n r) j_l(k_{n'} r) = \delta^K_{nn'} \frac{R^3}{2} [j_l(k_n R)]^2 \equiv \delta^K_{nn'} (C_{ln})^{-1} \quad , \tag{3}$$

where $C_{ln}$ represents the normalization coefficient (see Table A1) and $\delta^K_{ij}$ denotes the Kronecker delta symbol. The orthogonality of the radial and angular coordinates allow the complex harmonic coefficients in Equation 1 to be recovered by the following inversion formula,

$$\rho_{lmn} = \int_{V_R} d^3\mathbf{r}\, j_l(k_n r) Y^*_{lm}(\hat{\mathbf{r}}) \rho(\mathbf{r}) \quad , \tag{4}$$

where $V_R$ denotes an spherical integration region of radius $R$.

As we will show in the following sections, Equations 1 and 4 provide a convenient way of describing the redshift space density field. When we model dynamics, we will be interested in density fluctuations, $\rho(\mathbf{r}) = \bar{\rho}\,(1 + \delta(\mathbf{r}))$. A useful expression relating the harmonic coefficients of the density field to the fluctuation field is given by

$$\delta_{lmn} = \rho_{lmn}/\bar{\rho} - \sqrt{4\pi} R^3 \left( \frac{j_1(k_n R)}{k_n R} \right) \delta^K_{l0} \delta^K_{m0} \quad , \tag{5}$$

The second term in Equation 5 is the spherical transform of a constant which only contributes to the monopole ($l = 0$) harmonic (cf., Equation D14).

The direct observable in a redshift catalog is the *redshift* space density field and its harmonics denoted by $\rho_{lmn}^S$ [*]. In a flux-limited catalog, the density field is is sampled by a discrete set of galaxies. An estimate of the density field [†] is given by

$$\hat{\rho}(\mathbf{s}) = \sum_{i=1}^{N} \frac{1}{\phi(s_i)} \delta^{(3)}(\mathbf{s} - \mathbf{s}_i) \qquad (6)$$

where $\phi(s)$ is the selection function of the survey evaluated at the redshift of the galaxy (in a homogeneous Euclidean universe, the number of galaxies on a radial shell is $\propto r^2 \phi(r)$) and $N$ is the number of galaxies within the spherical region, $s < R$. [‡] If we substitute this estimator into Equation 4, we obtain the following expression for the harmonics of the redshift space density field,

$$\hat{\rho}_{lmn}^S = \sum_{i=1}^{N} \frac{1}{\phi(s_i)} j_l(k_n s_i) Y_{lm}^*(\hat{\mathbf{r}}_i) \qquad . \qquad (7)$$

This is the spherical analog of the Fourier coefficients used in analyses of the power spectrum (e.g., Fisher *et al.* 1993)

Finally, if we subtract the contribution due to the mean number density as given in Equation 5 from $\hat{\rho}_{lmn}^S$, we obtain an estimate of the harmonics of the fluctuation field in redshift space, $\hat{\delta}_{lmn}^S$. We have adopted the minimum variance estimator the mean density suggested by Davis & Huchra (1982):

$$\bar{\rho} = \frac{\sum_{i=1}^{N} g(s_i)}{\int\limits_0^\infty ds\, s^2 \phi(s) g(s)} \qquad , \qquad (8)$$

where the weighting function is given by

$$g(s) = \frac{1}{1 + 4\pi \bar{\rho}_1 J_3(20 h^{-1} \mathrm{Mpc})\, \phi(s)} \qquad , \qquad (9)$$

which, incidentally, is an example of a Wiener filter to be described in § 4. In Equation 9, we need an initial guess for the mean number density, $\bar{\rho}_1$, and an estimate of $J_3(r) = \int_0^r dx\, x^2 \xi(x)$. We adopt fiducial values appropriate for the 1.2 Jy *IRAS* sample: $\bar{\rho}_1 = 0.041\, (h\mathrm{Mpc})^{-3}$ and $J_3(20\, h^{-1}\mathrm{Mpc}) = 4680\, (h\mathrm{Mpc})^3$ (Fisher *et al.* 1994a). These values give $\bar{\rho} = 0.045 \pm 0.001$ for the *IRAS* 1.2 Jy survey.

## 3  REDSHIFT DISTORTION: RADIAL MODE COUPLING

In order to predict the peculiar velocity field of galaxies, we need an expression that relates the harmonics in redshift space to the harmonics in *real* space, $\delta_{lmn}^R$. Redshifts and distances are related by the transformation,

$$\mathbf{s} = \mathbf{r} + [\mathbf{v}(\mathbf{r}) - \mathbf{v}(\mathbf{0})] \cdot \hat{\mathbf{r}} \qquad (10)$$

where $\mathbf{v}(\mathbf{r})$ is the peculiar velocity field. In writing Equation 10, we have accounted for the motion of the observer; redshifts in this frame are referred to as Local Group (LG) redshifts. If peculiar velocities are small relative to the distance, then the redshift will be close to the actual distance. This motivates a perturbative approach, where redshift space is viewed as only slightly distorted from real space with the difference being important only to lowest order in the velocity field. This approach is valid within the context of linear theory where the density contrasts (and hence velocities) are relatively small and the transformation in Equation 10 is one-to-one.

This perturbative approach is quite convenient and allows the first order contribution to the redshift distortion to be calculated analytically; this derivation is given in Appendix D. Without going into the mathematical details, one can anticipate the structure of the distortion term. The key point is that peculiar velocities introduce only a *radial* distortion. Consequently, since we have expanded the density field in orthogonal radial and angular coordinates, then the effect of the distortion will be only to couple the radial modes while leaving the angular modes unaffected; the simplification of the distortion in this basis is the primary motivation for choosing spherical harmonics over plane waves.

It is convenient to describe the mode coupling by introducing a coupling matrix,

---

[*] We will adopt the use of super and subscripts $R$ and $S$ to denote quantities in real and redshift space respectively.
[†] Throughout this paper we denote estimates of quantities by carets over the actual quantity.
[‡] Note that although the radial selection, $\phi(r)$, is in real space, it is actually determined from redshift surveys. In practice, there is little difference between the *functional form* of $\phi(s)$ and $\phi(r)$.

$$\hat{\delta}^S_{lmn} = \sum_{n'} (\mathbf{Z}_l)_{nn'} \, \delta^R_{lmn'} \quad . \tag{11}$$

Note that the coupling matrix (for a sample of $4\pi$ coverage) only mixes radial modes with the same value of $l$. For samples with incomplete sky coverage there will be a more complicated coupling of angular modes at different $(l, m)$. In principle the analysis given below can be extended to this more general case but at the price of making the mathematics much more cumbersome. In practice, near full sky catalogues like those from extracted from the *IRAS* survey have almost complete coverage for $|b| > 5°$ and the extension of the method to incomplete sky coverage is not warranted for $l_{max} \lesssim 15$. (see § 6 below and the discussion in LFHSZ and FSL).

In linear theory, one can obtain an analytic expression for the coupling matrix; this derivation can be found in Appendix D. Schematically, the coupling matrix is the identity matrix (i.e,. the contribution from the undistorted real space harmonics) plus a distortion matrix which represents the mode-mode coupling in redshift space. This distortion matrix is independent of the power spectrum but it does depend both on the (known) selection function of the survey and on the (uncertain) quantity $\beta \equiv \Omega^{0.6}/b$ which dictates the amplitude of the velocity field in linear theory ($b$ is the linear bias parameter assumed throughout to be independent of scale).

The mode coupling in Equation 11 is a generic feature of the density field in redshift space and is not intrinsic to our choice of spherical coordinates. Indeed, the importance of a correct treatment of the mode-mode coupling in redshift space was emphasized by Zaroubi & Hoffman (1994) who derived an expression for the coupling matrix in a Cartesian representation using conventional Fourier analysis.

In a perfect galaxy catalogue with arbitrarily high sampling density, one could obtain an estimate of the real space harmonics by simply inverting the coupling matrix in Equation 11. Unfortunately, any estimate of the redshift space harmonics derived from an actual catalogue will necessarily be corrupted by stochastic noise arising from the discreteness of the galaxy distribution, commonly referred to as shot noise. In the presence of this noise, a straightforward inversion of the radial coupling matrix may amplify the shot noise and lead to an estimate of the real space harmonics which is far from optimal. The basic question which needs to be addressed is how to perform the inversion in such way that the statistical noise is minimized. There exists a vast literature on the subject of how to reconstruct a signal corrupted by noise, and there are a corresponding wide variety of methods which can be employed. In this paper, we have adopted perhaps the simpliest method, known as Wiener filtering, which is designed to minimize the variance between the recovered and true signal. In the next section we give a brief overview of the principles of Wiener filtering and then proceed to apply it to the inversion of Equation 11.

## 4 MINIMUM VARIANCE ESTIMATES OF COSMOLOGICAL FIELDS: WIENER RECONSTRUCTION

### 4.1 Concept of the Wiener Filter

This section gives a brief review of the WF technique; the reader is referred to LFHSZ, ZHFL, and Rybicki & Press (1992) for further details. Let us assume that we have a set of measurements, $\{d_\alpha\}$ ($\alpha = 1, 2, \ldots N$) which are a linear convolution of the true underlying signal, $s_\alpha$, plus a contribution from statistical noise, $\epsilon_\beta$, i.e.,

$$d_\alpha = \mathcal{R}_{\alpha\beta} \left[ s_\beta + \epsilon_\beta \right] \quad , \tag{12}$$

where $\mathcal{R}_{\alpha\beta}$ is the response or "point spread" function (summation convention assumed). Astronomical examples of the response function include "seeing" (e.g., Lucy 1974), the throughput of the pre-corrected optics of the Hubble Space Telescope, and the masked region (Zone of Avoidance) in the galaxy distribution (LFHSZ). In the present context it would be the radial coupling matrix discussed in the previous section. Notice that we have assumed that the statistical noise is present in the underlying field and therefore is convolved by the response function; this is the appropriate formulation for our reconstruction problem (where the shot noise is in the underlying albeit unknown real space density field) and differs from the case where the noise is due to measurement errors (e.g., see Bunn *et al.* 1994).

The WF is the *linear* combination of the observed data which is closest to the true signal in a minimum variance sense. More explicitly, the WF estimate is given by $s_\alpha(WF) = F_{\alpha\beta} \, d_\beta$ where the filter is chosen to minimize $\langle |s_\alpha(WF) - s_\alpha|^2 \rangle$. It is straightforward to show (see ZHFL for details) that the WF is given by

$$F_{\alpha\beta} = \langle s_\alpha d_\gamma \rangle \langle d_\gamma d^\dagger_\beta \rangle^{-1} \quad , \tag{13}$$

where

$$\begin{aligned} \langle s_\alpha d^\dagger_\beta \rangle &= \mathcal{R}_{\beta\gamma} \langle s_\alpha s^\dagger_\gamma \rangle \\ \langle d_\alpha d^\dagger_\beta \rangle &= \mathcal{R}_{\alpha\gamma} \mathcal{R}_{\beta\delta} \left[ \langle s_\gamma s^\dagger_\delta \rangle + \langle \epsilon_\gamma \epsilon^\dagger_\delta \rangle \right] \quad . \end{aligned} \tag{14}$$

In the above equations, we have assumed that the signal and noise are uncorrelated. From Equation 14, it is clear that in order to implement the WF one must construct a *prior* which depends on the variance of the signal and noise.

The dependence of the WF on the prior can be made clear by defining signal and noise matrices given by $S_{\alpha\beta} = \langle s_\alpha s_\beta^\dagger \rangle$ and $N_{\alpha\beta} = \langle \epsilon_\alpha \epsilon_\beta^\dagger \rangle$. With this notation, we can rewrite Equation 13 as

$$\mathbf{s}(WF) = \mathbf{S}\,[\mathbf{S}+\mathbf{N}]^{-1}\,\mathcal{R}^{-1}\,\mathbf{d} \quad . \tag{15}$$

Formulated in this way, we see that the purpose of the WF is to attenuate the contribution of low signal to noise ratio data and therefore regularize the inversion of the response function. The derivation of the WF given above follows from the sole requirement of minimum variance and requires only a model for the variance of the signal and noise. The WF can also be derived using the laws of conditional probability if the underlying distribution functions for the signal and noise are assumed to be Gaussian; in this more restrictive case, the WF estimate is, in addition to being the minimum variance estimate, also both the maximum a *posterior* estimate and the mean field (cf., LFHSZ, ZHFL). For Gaussian fields, the mean WF field can be supplemented with a realization of the expected scatter about the mean field to create a realization of the field; this is the heart of the "constrained realization" approach described in Hoffman & Ribak (1991; see also ZHFL).

As Rybicki & Press (1992) point out, the WF is in general a biased estimator of the mean field unless the field has zero mean; this is not a problem here since we will perform the filtering on the density fluctuation field which has, by construction, zero mean.

### 4.2 Reconstruction of the Density Field

In our specific case, the minimum variance solution for the real space harmonics of the density field is (in analogy with Equation 15 with $\mathbf{Z}_l$ playing the role of the response function) given by

$$\hat{\delta}^R_{lmn}(WF) = \sum_{n'n''} \left(\mathbf{S}_l\,[\mathbf{S}_l+\mathbf{N}_l]^{-1}\right)_{nn'} \left(\mathbf{Z}_l^{-1}\right)_{n'n''} \hat{\delta}^S_{lmn''} \quad , \tag{16}$$

where the signal and noise matrices are given by (cf., Appendix E)

$$(\mathbf{S}_l)_{nn'} = \frac{2}{\pi}\bar{\rho}^2 \int_0^\infty dk\,k^2 P(k) \int_0^R dr_1\,r_1^2 j_l(k_n r_1) j_l(k r_1) \int_0^R dr_2\,r_2^2 j_l(k_{n'} r_2) j_l(k r_2) \approx \bar{\rho}^2 P(k_n) C_{ln}^{-1} \delta^K_{nn'}$$

$$(\mathbf{N}_l)_{nn'} = \bar{\rho} \int_0^R dr\,r^2 \frac{j_l(k_n r) j_l(k_{n'} r)}{\phi(r)} \quad , \tag{17}$$

and the radial coupling matrix is

$$(\mathbf{Z}_l)_{nn'} = \delta^K_{nn'} - \beta \int_0^R dr\,r^2 j_l(k_n r) \left[ j_l''(k_{n'} r) + \left( \frac{j_l'(k_{n'} r)}{k_{n'} r} - \frac{\delta^K_{l1}}{3} \frac{1}{r} \int_0^R dx\,j_l(k_n' x) \right) \left( 2 + \frac{d\ln\phi(r)}{d\ln r} \right) \right] \quad . \tag{18}$$

To summarize the method so far: given a full sky redshift survey, one can compute the redshift space harmonics using the estimator in Equation 7. Next, for a given choice of $\beta$ and power spectrum (the *prior* information), one can compute both the radial coupling matrix $\mathbf{Z}_l$ and the expected signal and noise matrices in Equation 17. The real space harmonics can then be estimated from the redshift space harmonics using Equation 16. The WF in Equation 16 will attenuate the harmonics whose expected signal falls below the expected noise given by Equation 17.

### 4.3 Reconstruction of the Peculiar Velocity Field

The WF inversion of the coupling matrix provides us with estimates of the harmonics in real space. Given the density field, we can in linear theory compute the peculiar velocity field. Once more the spherical harmonic basis is convenient, and will allow the harmonics of the density and velocity fields to be related by a linear transformation.

The radial peculiar velocity field, being a scalar quantity like the density, can also be decomposed in its harmonics,

$$v_r(\mathbf{r}) = \sum_{lmn} C_{ln}\,v_{lmn}\,j_l(k_n r) Y_{lm}(\hat{\mathbf{r}}) \quad . \tag{19}$$

Moreover, in linear theory the radial velocity, due to the inhomogeneities within $r < R$, can be expressed directly in terms of the harmonics of the density field[§] (cf., Appendix C, FSL, Regös & Szalay 1989),

---

[§] We have chosen to work in velocity units with the Hubble constant set to unity, ie, $H_0 = 1$.

$$v_r(\mathbf{r}) = \beta \sum_{lmn} C_{ln} \delta^R_{lmn} \frac{j'_l(k_n r)}{k_n} Y_{lm}(\hat{\mathbf{r}}) \qquad . \tag{20}$$

Using Equations 19 and 20 from Appendix C, one can show that the velocity and density harmonics are related by a matrix equation, i.e.,

$$v_{lmn} = \beta \sum_{n'} (\mathbf{\Xi}_l)_{nn'} \delta^R_{lmn'} \qquad , \tag{21}$$

where

$$(\mathbf{\Xi}_l)_{nn'} = \frac{1}{k_{n'}} C_{ln'} \int_0^R dr\, r^2 j_l(k_n r) j'_l(k_{n'} r) \qquad . \tag{22}$$

The linear relationship between the density and velocity harmonics is particularly convenient in the context of the Wiener filter since it means that the minimum variance mean field velocity harmonics, $v_{lmn}(WF)$, are related to the minimum variance density harmonics by the matrix $\mathbf{\Xi}_l$, i.e.,

$$\begin{aligned} v_{lmn}(WF) &= \beta \sum_{n'} (\mathbf{\Xi}_l)_{nn'} \delta^R_{lmn'}(WF) \\ &= \beta \sum_{n'n''n'''} (\mathbf{\Xi}_l)_{nn'} \left(\mathbf{S}_l [\mathbf{S}_l + \mathbf{N}_l]^{-1}\right)_{n'n''} \left(\mathbf{Z}_l^{-1}\right)_{n''n'''} \hat{\delta}^S_{lmn'''} \qquad . \end{aligned} \tag{23}$$

Therefore, having derived the minimum variance estimates of the density harmonics, we can also obtain the minimum variance harmonics of the radial velocity field by a simple multiplication of the matrix $\mathbf{\Xi}_l$.

The harmonics of the transverse component of the velocity field, $\mathbf{v}_\perp(\mathbf{r})$, can, in linear theory, also be related to the those of the density field. The expressions for the transverse velocity field are slightly more complicated than for the radial case and are derived in Appendix C.2.

### 4.4 Reconstruction of the Potential Field

In Appendix C, we show that the harmonics of the potential field, $\psi(\mathbf{r})$ are related to those of density field by $\psi_{lmn} = -3/2\Omega/k_n^2 \delta_{lmn}$. Once again, the minimum variance estimates of the potential harmonics are related to the WF harmonics of the density field by a linear transformation,

$$\begin{aligned} \psi_{lmn}(WF) &= \Omega \sum_{n'} (\mathbf{\Phi}_l)_{nn'} \delta^R_{lmn'}(WF) \\ &= \Omega \sum_{n'n''n'''} (\mathbf{\Phi}_l)_{nn'} \left(\mathbf{S}_l [\mathbf{S}_l + \mathbf{N}_l]^{-1}\right)_{n'n''} \left(\mathbf{Z}_l^{-1}\right)_{n''n'''} \hat{\delta}^S_{lmn'''} \qquad , \end{aligned} \tag{24}$$

where

$$(\mathbf{\Phi}_l)_{nn'} = -\frac{3}{2} \frac{1}{k_n^2} \delta^K_{nn'} \qquad . \tag{25}$$

### 4.5 Choice of Reference Frame

Our formulation for the redshift distortion takes into account the motion of the observer $\mathbf{v}(\mathbf{0})$ (starting from Equation 10), and as such the choice of reference frame is arbitrary. However, the modeling in linear theory involves a Taylor expansion out to first order in $\Delta \mathbf{v} = (\mathbf{v}(\mathbf{r}) - \mathbf{v}(\mathbf{0}))$ (cf., Equations D4 and D5), so working in the frame in which this velocity difference is small will yield better reconstruction. From measurements of peculiar velocities in the local universe it is known that a sphere of radius $R_1 \sim 3000$ km s$^{-1}$ is sharing the motion of the Local Group with respect to the CMB (Sandage 1986; Brown & Peebles 1987; Faber & Burstein 1988), although the value of this "coherence length" is rather uncertain (Górski *et al.* 1989). Hence, out to $R_1$ it is better to correct for the redshift distortion in the LG frame. However, at larger distances (say $R_2 \sim 10,000$ km s$^{-1}$) galaxy motions are no longer strongly correlated with the motion of the LG and $\Delta \mathbf{v}(\mathbf{r})$ is smaller in the microwave background (CMB) frame (where $\mathbf{v}(\mathbf{0})$ is then omitted from Equation 10). In principle, other frames are also possible; for example, one can attempt to model the transition from the LG to CMB frame in order to minimize $\Delta \mathbf{v}$ as a function of distance (e.g., Strauss et al. 1992a). Since we will mainly be interested in reconstructions of the local velocity field, we have chosen to work in the LG frame.

Although the observed harmonics are computed using LG redshifts, the reconstructed harmonics will be in real space after the inversion of the radial coupling matrix. Thus, the radial velocity field reconstruction from Equation 23 will be relative to the rest frame defined pure Hubble flow, i.e. in the CMB frame. However, we can put the reconstructed velocities into the LG frame by subtracting out the reconstructed observer's velocity (cf., Equation C12). Comparing velocities in the LG frame helps to avoid systematic errors that might arise from our neglect of the gravitational contribution from galaxies outside $r > R$. The dominant effect of the "missing power" due to the material with $r > R$ will be to induce a dipole or bulk flow within $r < R$ when measured in the CMB frame (Juszkiewicz, Vittorio, & Wyse 1990; Lahav, Kaiser, & Hoffman 1990; YSDH; ND). The advantage of the LG frame is that it measures *relative* velocities.

Kaiser (1987) pointed out that spurious motions could be inferred from a redshift galaxy distribution if the density field was computed in redshift space. He gave the hypothetical example of a homogeneous universe with no peculiar velocities but with an observer viewing the galaxies in redshift space from a moving rocket. In redshift space the observer will infer a dipole distribution due to the Doppler shift induced the rocket's motion despite the fact that the galaxies are uniformly distributed throughout space and have zero peculiar velocities. In a realistic catalogue this "rocket effect" limits the useful information which can be gleaned from the Local Group dipole (Strauss *et al.* 1992a). One powerful advantage of working in the spherical harmonic basis is that the rocket effect is isolated in the dipole or $l = 1$ harmonic. In our analysis the rocket effect can be modeled self-consistently taking into account the distortion of the density field in redshift space (cf., Appendix D).

## 5   THE ALGORITHM AND $N$-BODY TESTS

The integrals which appear in the radial coupling matrix, $\mathbf{Z}_l$, and the velocity-density coupling matrix, $\mathbf{\Xi}_l$ can be computed numerically for each $l$. The spherical Bessel functions are evaluated using stable forms of the recursion relations (e.g., Press *et al.* 1992) while the numerical quadrature is done with NAG (1980) routines designed to handle oscillatory integrands. Inversion of the matrices ($\mathbf{Z}_l$ and $[\mathbf{S}_l + \mathbf{N}_l]$) were carried out using the method of Singular Value Decomposition (SVD) (cf., Press *et al.* 1992, LFHSZ, ZHFL). In practice the matrix inversions were very stable over the range of $l$ considered ($l_{max} = 15$) and consequently the SVD approach gave identical results to standard matrix inversion.

In order to test the whole reconstruction procedure, we applied the method to mock galaxy catalogues extracted from a numerical simulation of a CDM universe. The simulations are those used by Górski *et al.* (1989), Frenk *et al.* (1990), and Davis, Strauss, & Yahil (1991), and are of a standard biased CDM universe ($h = 0.5, \Omega = 1, \Lambda = 0, n = 1$). The output time of the simulation corresponds an *rms* amplitude of mass fluctuations in an 8 $h^{-1}$Mpc sphere of $\sigma_8 = 0.62$. The points selected as galaxies, however, are not biased; they do trace the mass, i.e., $b = 1$ and $\beta = \Omega^{0.6}/b$ is unity. Mock catalogues were constructed by choosing particles from the simulation that match the observational constraints of the Local Group as described in Górski *et al.* .

The particles in the simulations were assigned luminosities based on the luminosity function of the 1.2 Jy *IRAS* sample (Fisher 1992) and a series of flux-limited catalogues were extracted. These mock catalogues closely resemble the 1.2 Jy *IRAS* survey in sampling density. The velocity field of the simulations has been convolved with a Gaussian window of width 1 $h^{-1}$Mpc in order to reduce the pair velocity dispersion on small scales to value close to that observed, $\sigma(1h^{-1}\text{Mpc}) \sim 317$ km s$^{-1}$ (Fisher *et al.* 1994b). In this paper, we assume that the sky coverage is complete over $4\pi$ steradians. It is straightforward, at least conceptually, to generalize the method to the case of incomplete sky coverage (e.g., using the mask inversion technique presented in LFHSZ). However, modeling the incomplete sky coverage greatly increases the computational complexity of the reconstruction. Moreover, our final goal will be to apply the method to near full-sky catalogues such as the 1.2 Jy *IRAS* survey which covers 87.5% of the sky. For such samples the statistical corrections for the missing sky are small (LFHSZ) and it is adequate to smoothly interpolate data in the missing regions (cf., YSDH).

Figure 1 shows the effect of the redshift distortion and Wiener filtering in the reconstruction process. We have reconstructed the peculiar velocity field from the mock catalogues by performing the harmonic expansion for all galaxies within a sphere of radius 180 $h^{-1}$Mpc using the estimator given in Equation 7. The harmonics are computed up to $l_{max} = 15$ in angle and for $k_n R \leq 100$ in the radial modes. The peculiar velocities are reconstructed within $r < 60$ $h^{-1}$Mpc. In each of the four panels in this Figure, we plot the reconstructed radial velocities versus the true (nonlinear) radial velocities as given in the $N$-body simulation.

In order to illustrate the effects of the redshift distortion and shot noise separately, the upper left panel shows the reconstructed velocities if one neglects altogether both the effects of redshift distortion and shot noise in the harmonics, i.e., treating the WF and coupling matrices in Equation 23 as the identity matrix; this is equivalent to solving the standard Poisson's equation by treating the redshifts as actual distances and neglecting the effects of shot noise. Although the correlation between the reconstructed and true velocities is fairly good, the slope differs substantially from unity. This is a reflection of the enhanced density contrasts in the redshift space.

In the upper right hand panel of Figure 1, we show the reconstructed velocities when again redshift distortions are neglected but the statistical noise is lessened by the application of the Wiener filter. The Wiener filter reduces the scatter in the reconstructed velocities and improves the slope by smoothing the density fluctuations, yet the velocities are still overpredicted due to the neglect of the redshift distortions in the density field. In the lower left hand panel, we show the reconstructed velocities when the effects of redshift distortion have been modeled (by taking the inverse of the radial coupling matrix) but with no Wiener filtering applied. In this case, there is an improvement over working purely in redshift space but the velocities are still slightly overestimated.

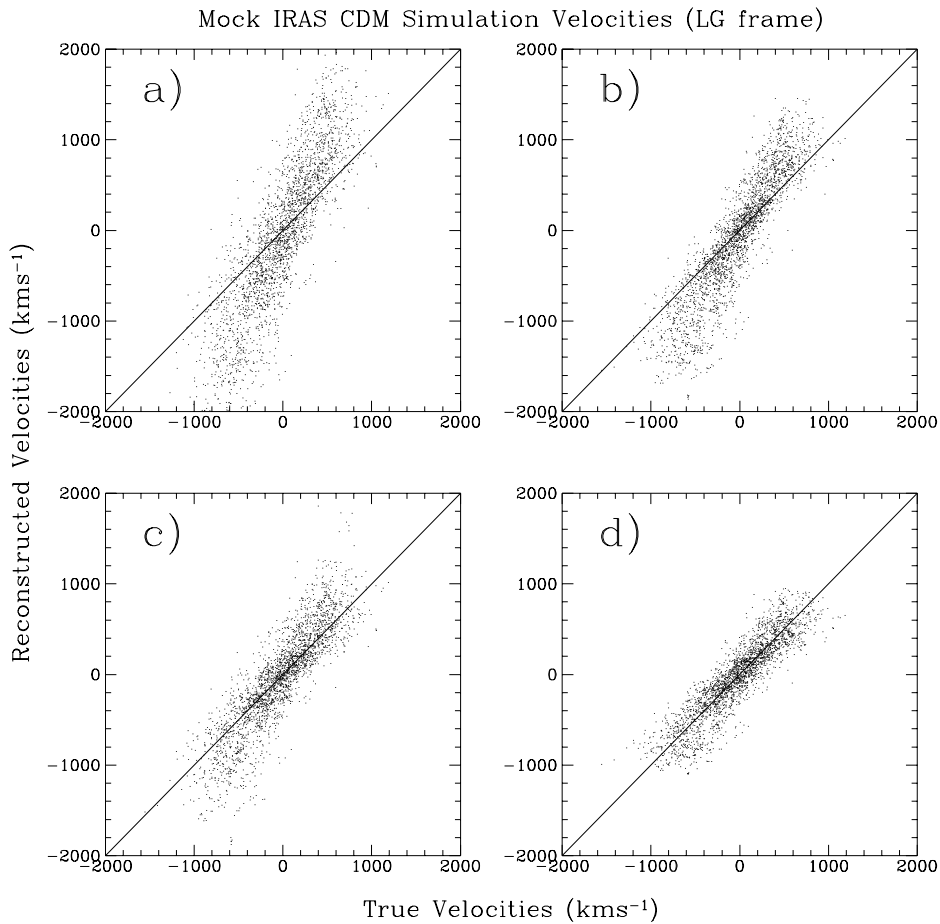

**Figure 1.** Radial velocity field reconstructions of the mock *IRAS* CDM catalogues. The reconstructions are performed within a sphere of radius 180 $h^{-1}$ Mpc and, the velocities are shown for galaxies within $r < 60$ $h^{-1}$ Mpc. a.) Pure redshift space with no Wiener filter applied. b.) Here the harmonics have been smoothed by the Wiener filter, but no dynamical correction for the redshift distortion has been made. c.) Correction for redshift distortion has been performed but using the unsmoothed raw redshift harmonics. d.) Full reconstruction correcting for the redshift distortion and including the Wiener filtering of the harmonics.

Finally, in the lower right hand panel we show the complete reconstruction where both the redshift distortions correction and the Wiener filter have been applied. The full reconstruction matches the true velocities quite well with no apparent gross systematic effects. In fact, the *rms* velocity difference between the reconstructed and true velocities is $\sim 190$ km s$^{-1}$ (see Table 1) and suggests that the method is performing quite well in moderate to low density regions since the *rms* pairwise velocity dispersion of the $N$-body model in high density nonlinear regimes is $\sim 350$ km s$^{-1}$. A simple least squares fit to the data in Figure 1*d* yields a slope of 0.94 indicating no large systematic bias in the reconstruction. The value of slope and *rms* errors for the various reconstructions in Figure 1 are given in Table 1.

Figures 1*c* and 1*d* highlight two competing effects in our reconstruction. Since we only assume linear theory for the dynamics, the reconstructed velocities are slightly larger than the true velocities (as indicated by a slope of 1.18 in Figure 1*c*, see Table I). This effect of velocity overestimation when using only linear theory can be seen from the second order approximation to the spherical infall model (e.g., Lightman & Schechter 1990) or from an empirical fit to the Zeldovich approximation (Nusser *et al.* 1991).

This effect can be understood qualitatively as follows. The nonlinear peculiar velocity field is more similar to the linear velocity field than the nonlinear density field is to the linear density field, simply because the velocity field is influenced by structure on larger scales which are more likely to be still in the linear regime. Thus, near a site of nonlinear collapse the velocity field will be amplified by a factor which is typically much smaller than the density field. Hence by applying linear theory (where the Fourier amplitudes of the velocity and density are proportional to one another) to an underlying nonlinear density field we will systematically *over*estimate the peculiar velocity. When the WF is applied (Figure 1*d*) high density regions and hence high velocities are attenuated, bringing the slope down to 0.94 (Table I), nearer to unity. This "cooperation" where the dynamical and statistical effects tend to almost cancel each other is as much of a "cosmic coincidence" as the similarity between the *IRAS* galaxy-galaxy correlation length ($r_o \sim 4\ h^{-1}$Mpc, which defines a nonlinear dynamical scale) and the mean separation distance between *IRAS* galaxies ($\bar{\rho}^{-1/3} \sim 3\ h^{-1}$Mpc, which defines a statistical shot noise scale).

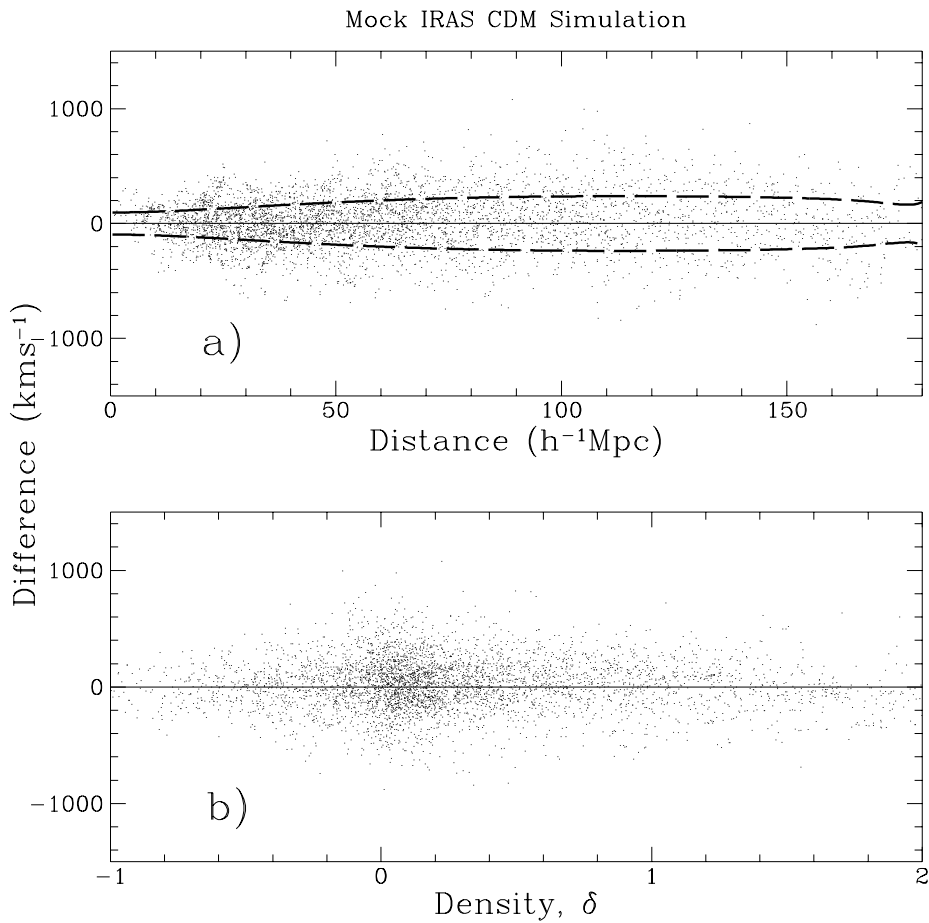

**Figure 2.** a.) The difference between the reconstructed and true nonlinear peculiar velocities (in Local Group frame) versus distance for the mock *IRAS* CDM simulation. The dashed curves show the expected scatter in the reconstructed velocities (see Appendix F). b.) Same velocity differences in a) shown as a function of the reconstructed density contrast.

In order to check for possible systematic biases in the method, we show the difference between the reconstructed and true velocities as a function of both distance and density in Figure 2. In the upper panel we plot the velocity difference as a function of the distance from the central observer in the simulation. The dark dashed curves represent the expected scatter in the velocities due to the Wiener filtering (cf., Appendix F). The decrease in the expected scatter near $r = 0$ and $r = R$ is consequence of the imposed boundary conditions. There does not appear to be any systematic offset in the mean difference as a function of distance which might, for example, arise if the monopole of the density field was in error. In the lower plot of Figure 2 we plot the same velocity differences but now as a function of the reconstructed density at the position of each particle. Once again no systematic trend is evident and the mean difference is consistent with zero over the entire range of densities probed.

## 6  APPLICATION TO THE 1.2 JY *IRAS* REDSHIFT SURVEY

In this section, we present a preliminary application of the reconstruction technique to the sample of 5313 galaxies in the 1.2 Jy *IRAS* survey (Fisher 1992; Strauss *et al.* 1992*b*). The formalism we have discussed is limited to samples with full $4\pi$ coverage. The 1.2 Jy survey covers 87.6% of sky with the incomplete regions being dominated by the 8.7 % of the sky with $|b| < 5°$. In principle the method can be extended to explicitly account for the incomplete sky coverage. We adopt the simplier and more expedient approach of smoothly interpolating the redshift distribution over the missing areas using the method described in YSDH. In Lahav *et al.* 1994, we examined the validity of this interpolation and found its effect on the computed harmonics to be negligible for $l \lesssim 15$ for the geometry of the 1.2 Jy survey. We therefore feel comfortable with using the interpolated catalogue (and simplified formalism) for our preliminary velocity reconstruction.

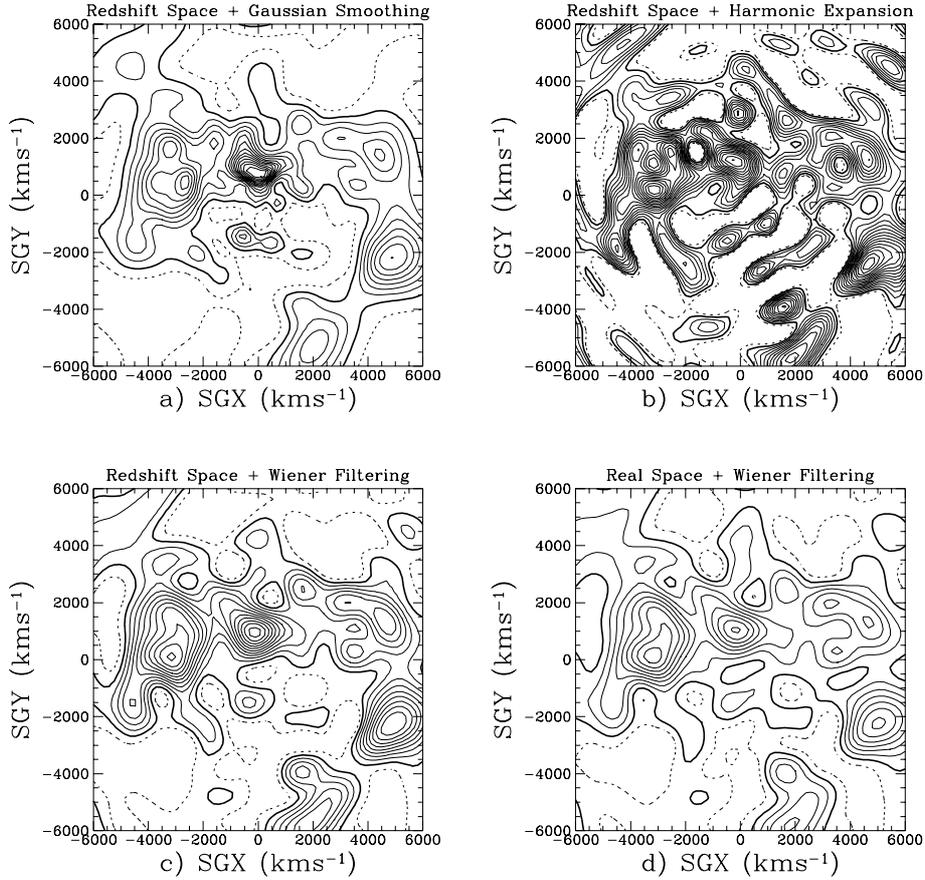

**Figure 3.** Reconstructions of the 1.2 Jy *IRAS* density field in the Supergalactic Plane. All contours are spaced at $\Delta\delta = 0.5$ with solid (dashed) lines denoting positive (negative) contours. The heavy solid contour corresponds to $\delta = 0$. a.) Raw redshift data smoothed with a Gaussian kernel of width proportional to the local mean interparticle separation. b.) Redshift space density field expanded in harmonics with $l_{max} = 15$ and $k_n R < 100$ but no additional smoothing. c.) Same as in b) but smoothed by the Wiener filter. d.) Same as in c) but with the harmonics corrected for redshift distortion (assuming $\beta = 1$).

In order to apply the WF algorithm, we need a model for the prior and this in turn depends on the power spectrum (in the actual WF), and on $\beta = \Omega^{0.6}/b$ (in the coupling matrix). Fortunately, the shape and amplitude of the power spectrum has been relatively well determined for *IRAS* galaxies. The power spectrum is well described (phenomenologically) on scales $\lesssim 200\ h^{-1}$Mpc by a CDM power spectrum with shape parameter $\Gamma$ (cf. Efstathiou, Bond, & White 1992) in the range $\Gamma \simeq 0.2-0.3$ (Fisher *et al.* 1993; Feldman, Kaiser, & Peacock 1993). The normalization of the power spectrum is conventionally specified by the variance of the galaxy counts in spheres of 8 $h^{-1}$Mpc, $\sigma_8$. Fisher *et al.* (1994a) used the projection of the redshift space correlation function to deduce the value, $\sigma_8 = 0.69 \pm 0.04$ in *real* space. The value of $\beta$ for *IRAS* is more uncertain; we have chosen to adopt the value found by FSL, $\beta = 1.0 \pm 0.3$, in their analysis based on redshift distortions in the spherical harmonic power spectrum. In what follows, we have adopted the WF prior given by $\Gamma = 0.2$ and $\sigma_8 = 0.7$ with $\beta = 1$. We have also performed the reconstruction of the *IRAS* velocities with a standard CDM ($\Gamma = 0.5$) prior. Since the standard CDM model has less large scale power than the $\Gamma = 0.2$ model, the WF smooths more on large scales and therefore the reconstructed velocities tend to be smaller; the overall difference is however small with $\langle \Delta v^2 \rangle^{1/2} \lesssim 50$ km s$^{-1}$.

Figure 3 shows several reconstructions of the *IRAS* density field in the Supergalactic Plane (SGP). The harmonics were reconstructed within a sphere of radius $R = 20,000$ km s$^{-1}$ with the angular modes limited to $l_{max} = 15$ and the radial modes restricted to $k_n R < 100$. For comparison, Figure 3a shows the redshift space density field derived by smoothing the raw data with a Gaussian window of width, $\sigma_s$, proportional to the mean interparticle separation. ($\sigma_s = 436$ km s$^{-1}$ at $r = 4000$ km s$^{-1}$, $\sigma_s = 626$ km s$^{-1}$ at 6000 km s$^{-1}$, and $\sigma_s = 1130$ km s$^{-1}$ at 10,000 km s$^{-1}$). The contours are spaced at $\Delta\delta = 0.5$ with solid (dashed) lines denoting positive (negative) contours. The large overdensity at SGX$\sim -3500$ km s$^{-1}$ and $SGY \sim 0$ km s$^{-1}$ is the Hydra-Centaurus supercluster complex, commonly referred to as the Great Attractor (GA). On the opposite side of the sky, the extended overdensity centered at (SGX$\sim 5000$ km s$^{-1}$, $SGY \sim -2000$ km s$^{-1}$) is the Perseus-Pisces (P-P) supercluster. Virgo and Ursa-Major are unresolved at this smoothing and appear as the single large overdensity at SGX$\sim 0$ and $SGY \sim 1000$ km s$^{-1}$. Figure 3*b* shows the unsmoothed harmonic expansion of the redshift space density field as given by Equation 1 with no correction for redshift distortion or Wiener filtering; the large statistical noise in this reconstruction demonstrates how strongly the harmonics can be corrupted by shot noise.

Figure 3*c* shows the redshift space density in panel *b* after the WF has been applied. As expected, the WF removes the noisy, high frequency harmonics from the map. The smoothing performed by the WF is variable and increases with distance. Comparison of Figure 3*c* with Figure 3*a*, shows that the WF is very similar to Gaussian smoothing with a dispersion proportional to the local mean interparticle separation. Figure 3*d* shows the reconstructed real space density field, $\delta(\mathbf{r})$, obtained from the density field in panel b) after correcting for both redshift distortion (assuming $\beta = 1$) and applying the WF. The correction of the redshift distortion has two effects. First, the overall amplitude of the density contrasts are reduced. Second, structures tend to become more spherical in real space as the compression parallel to line of sight induced by large scale streaming motions is lessened.

Figure 4*b* shows the potential reconstructed from the real space density field (shown again for comparison in Figure 4*a*). The contours are drawn at $\Delta\phi/c^2 = 0.05 \times 10^{-6}$ with the solid (dashed) contours representing positive (negative) values. The heavy contour corresponds to zero. The potential is much smoother than the density field (which via the Poisson's equation is obtained by twice differentiating the potential field). The GA and P-P superclusters dominate the potential field. The position of the potential maximum of the GA in our reconstruction is SGX$\sim -3500$ km s$^{-1}$, SGY$\sim 1500$ km s$^{-1}$, and SGZ$\sim 0$ km s$^{-1}$, which is somewhat nearer than that inferred by Lynden-Bell *et al.* (1988) (SGX = 4200 km s$^{-1}$, SGY= 760 km s$^{-1}$, SGZ = -690 km s$^{-1}$) and from the POTENT algorithm (Dekel, private communication). Of course the position of the extrema is dependent on the amount of smoothing which differs in these three methods. Note, however, that the peak of the GA in potential is shifted from the peak of the GA in number density. The shift in density and potential maxima could be due to the movement of density peaks during nonlinear evolution, but again the degree of displacement is a function of the smoothing.

The reconstructed radial velocity field ($\beta = 1$) is shown in Figure 4*c*. Positive (outflowing) radial velocities are shown as solid dots while negative radial velocities are denoted by open dots. The velocities shown here are in the CMB frame. Although it is more accurate to perform quantitative comparisons in the LG frame, it is easier to see the nature of the velocity field in the CMB frame. The general sense of the reconstructed radial velocity field is strong outflow towards both the GA and P-P superclusters and flow out of the Local Void (at SXY$\sim -2000$ km s$^{-1}$) along the positive SGY direction.

The reconstructed three dimensional peculiar velocity field is shown in Figure 4*d* and is obtained from the gradient of the potential field (details of this calculation are given in Appendix C.2). Again, the flow is dominated by the GA and P-P. The reconstruction shows a backside infall to the GA (in the CMB frame) of about $400 \pm 200$ km s$^{-1}$ (where the scatter is taken from Figure 2 at 40 $h^{-1}$Mpc). While at face value this supports the claim for a backside infall to the GA (Dressler & Faber 1990), we note that our velocity reconstruction is only due to the matter represented by galaxies within 20,000 km s$^{-1}$ and much more weight is given by the WF to the galaxies (as sources of gravity) nearby, within say 6000 km s$^{-1}$. This issue of backside infall and other cosmographical studies from the *IRAS* 1.2 Jy survey will be presented elsewhere.

# 7 DISCUSSION

## 7.1 Summary of the Method

To summarize, the key steps of this new reconstruction method are:

- **Expansion** in orthogonal functions (Equations 1 and 5) which satisfy Poisson's equation,

$$\delta(\mathbf{r}) = \sum_{lmn} C_{ln}\, \delta^S_{lmn}\, j_l(k_n r) Y_{lm}(\hat{\mathbf{r}}) \quad , \tag{26}$$

- **Inversion** of the coupling matrix to make a dynamical correction for redshift distortion assuming linear theory and a value of $\beta = \Omega^{0.6}/b$, (Equation 16)

$$\sum_{n'} \left(\mathbf{Z}_l^{-1}\right)_{nn'} \hat{\delta}^S_{lmn'} \quad , \tag{27}$$

- **Estimation** of the real space harmonics and suppression of statistical noise with the Wiener filter (Equation 16) for a chosen prior (i.e., a given power spectrum),

$$\hat{\delta}^R_{lmn}(WF) = \sum_{n'n''} \left(\mathbf{S}_l\left[\mathbf{S}_l + \mathbf{N}_l\right]^{-1}\right)_{nn'} \left(\mathbf{Z}_l^{-1}\right)_{n'n''} \hat{\delta}^S_{lmn''} \quad , \tag{28}$$

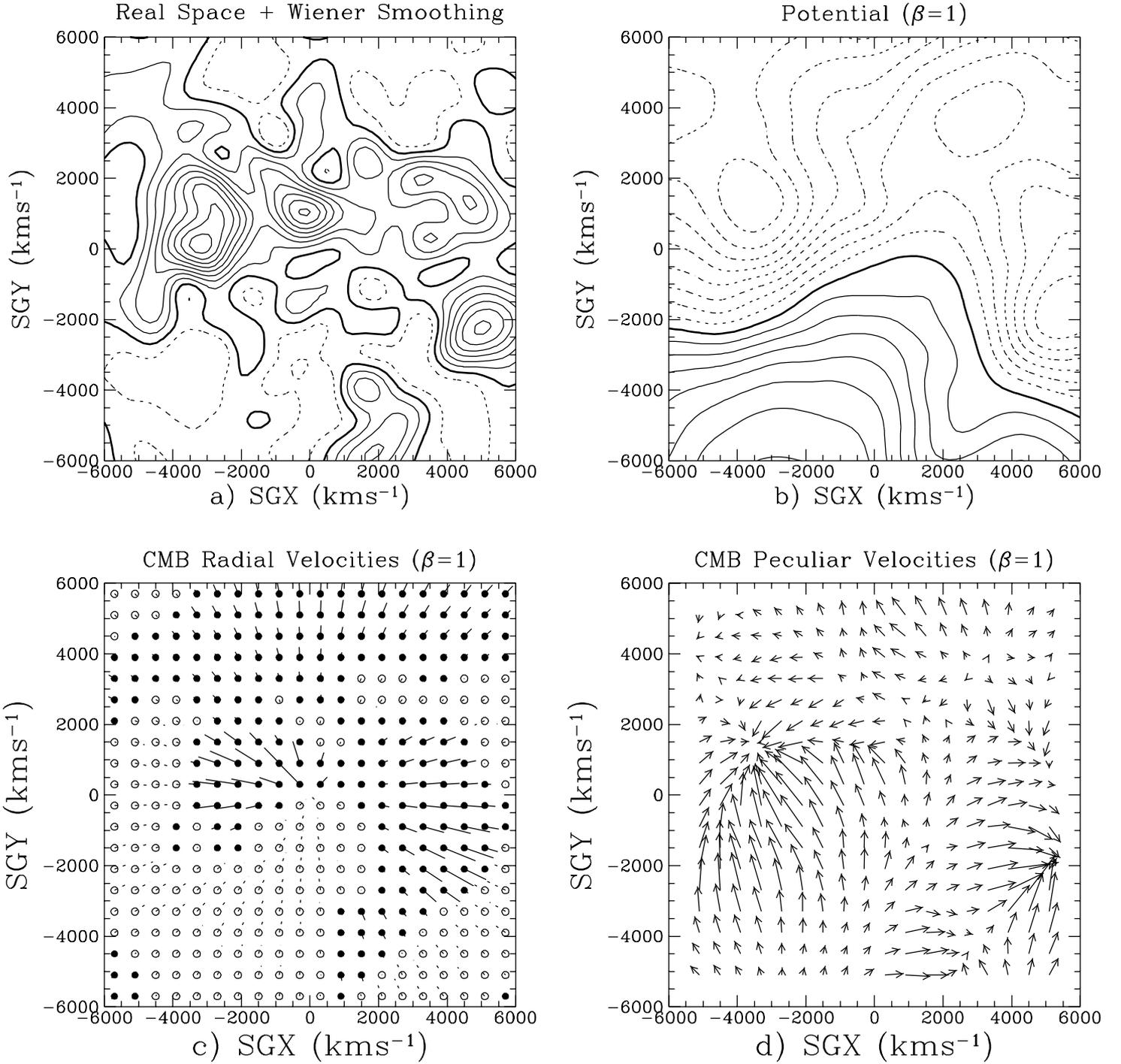

**Figure 4.** a.) The reconstructed real space density field for the 1.2 Jy *IRAS* sample in the Supergalactic Plane (SGP) as in Figure 3a. b.) Reconstructed dimensionless gravitational potential field, $\phi(\mathbf{r})/c^2$ from the 1.2 Jy survey for $\beta = 1$. Contours are spaced at $\Delta\phi/c^2 = 0.05 \times 10^{-6}$. Solid (dashed) contours denote positive (negative) values with the heavy contour representing $\phi = 0$. c.) Reconstructed radial velocity field. Closed (open) dots represent positive (negative) velocities. d.) Reconstructed three dimensional peculiar velocity field.

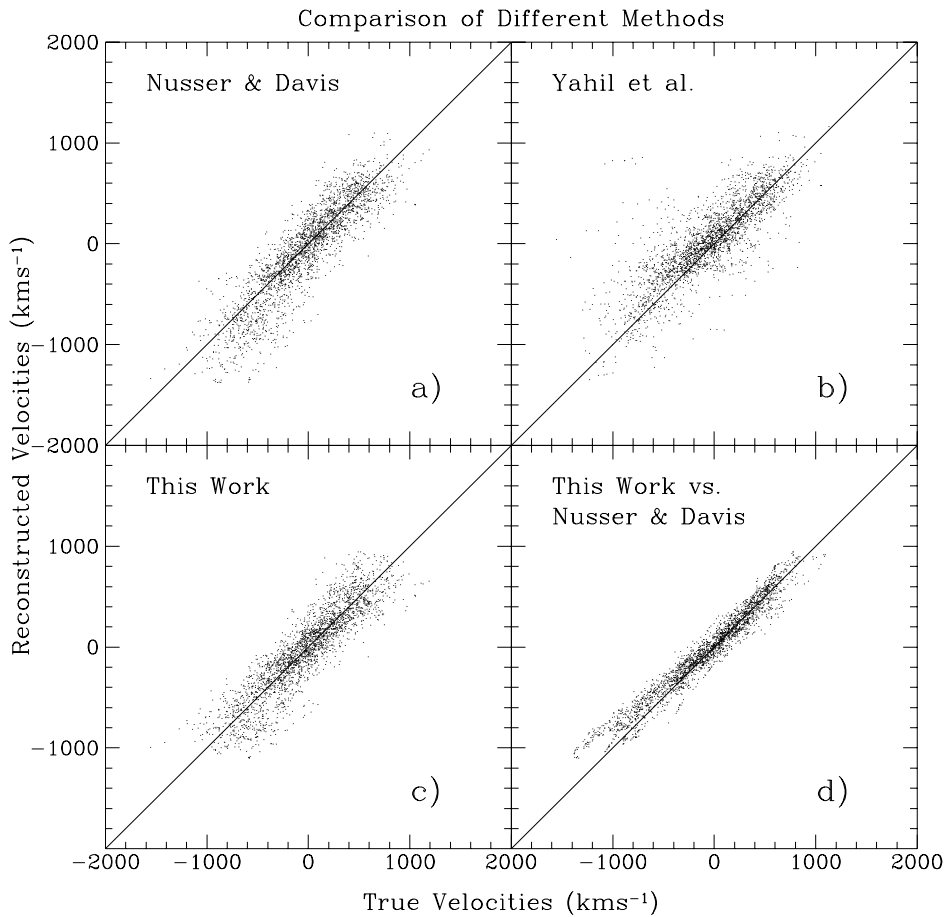

**Figure 5.** Comparison of different reconstruction techniques for the Mock IRAS CDM galaxies with distances less than 60 $h^{-1}$Mpc. a.) Velocities reconstructed using the technique of Nusser & Davis (1994) b.) Velocities reconstructed using the technique of Yahil *et al.* (1990). c.) Velocities reconstructed using the Wiener filter algorithm described in the text. d.) The Wiener method velocities (y-axis) versus the those reconstructed using the Nusser & Davis (x-axis) method.

- **Prediction** of the velocity field, either radial (Equation 23)

$$v_{lmn}(WF) = \beta \sum_{n'} (\Xi_l)_{nn'}\, \delta^R_{lmn'}(WF) \qquad , \tag{29}$$

or transverse (Equation C13), or the potential field (Equation 24) due to the mass distribution represented by galaxies inside the spherical volume.

We emphasize that the method is *non*-iterative. It provides a non-parametric and minimum variance estimates of the density, velocity, and potential fields which are related by simple linear transformations.

## 7.2 Comparison with Previous Work

Previous reconstruction methods for the peculiar velocity field from all-sky redshift surveys can roughly be divided into two approaches. The first technique applied was based on an iterative solution to the equations of linear theory. This approach was pioneered by YSDH) (see also Yahil 1988; Strauss & Davis 1988). This technique works by first solving for the gravity field in redshift space and then using it to derive an estimate for the peculiar velocities for a given value of $\beta$. These velocities then allow the redshifts to be corrected and provide an updated set of distance estimates. The whole process is repeated until the distance estimates converge. Variants of this technique have been successfully applied to *IRAS* selected galaxy catalogues (YSDH; Kaiser *et al.* 1991), optical catalogs (Hudson 1994) as well as *IRAS*/optical hybrid samples (Freudling *et al.* 1994).

The second set of methods are inherently non-iterative. These methods rely on a self-consistent formulation of the linear theory equations in redshift space; the reconstruction technique outlined in this paper is an example as is the method of ND.

Our method differs from ND in the way in which shot noise is treated; ND adopt a reasonable (albeit somewhat ad-hoc) smoothing of the density field prior to reconstruction, while the WF performs the smoothing in a natural way based on prior information. A more technical difference concerns the method used to expand the density field; ND looked at the angular harmonics on chosen shells and performed a spline fit to the radial trend whereas we expand the density field in orthogonal radial basis functions. The method of ND involves solving a differential equation corresponding to a modified Possion equation, while our method is essentially an algebraic solution to this equation in transform space.

Figure 5 is a comparison of the reconstructed peculiar velocity field of the mock *IRAS* CDM catalogue described in § 5 using three diferent methods: that of the WF presented here, YSDH, and ND (kindly provided to us by Michael Strauss and Adi Nusser). It is encouraging that the different methods agree as well as they do. Linear theory seems to be fairly robust to the algorithm which implements it, and while different smoothing methods may optimize the solutions, the recovered velocity field does not appear to be terribly sensitive to the exact smoothing technique employed. The comparison gives hope that the *IRAS* density and velocity fields can be reliably estimated out to about 6000 km s$^{-1}$. At larger distances, the sampling in the *IRAS* surveys becomes increasing dilute and one would expect the difference between methods to become more pronounced as regularization in the form of smoothing becomes more important.

### 7.3 Future Work

The WF reconstruction appears to perform very well and is ideally suited for noisy full-sky redshift surveys. The framework presented in this paper can, however, be extended in a number of ways. For example, the choice of Bessel functions in the expansion was motivated by simplicity in the dynamical equations. Other basis functions may provide much better resolution with fewer coefficients despite being mathematically more cumbersome. Second, one may envision performing the reconstruction within the Bayesian framework for a more realistic prior model. One natural extension would be to choose the prior of the underlying fluctuations as a log-normal distribution (Sheth 1994) which also ensures positivity of the reconstructed density field. A more realistic model for the shot noise would allow for spatial variations associated with peaks and voids. The WF could be modified to account for thermal noise in the nonlinear velocity field or else the redshift data might be "pre-processed" by collapsing the Fingers of God to remove any potential problems caused by nonlinear regions. Finally, one might relax the assumption of minimum variance and aim at designing a filter which preserves the moments of the density field (Yahil *et al.* 1994, in preparation). It is encouraging, however, that the simple implementation of the WF used in our analysis performs as well as it does. The WF reconstruction has the advantage over previous methods in that both the dynamics *and* the smoothing are performed in a natural way which incorporates our prior knowledge of large scale structure. We believe the method provides both an elegant and unified framework for describing cosmological fields.

### Acknowledgements

We would especially like to thank Adi Nusser for many useful and stimulating discussions and for helping with the comparisons between the methods. We thank Michael Strauss for critiquing an earlier draft and for providing the velocity predictions based on the Yahil *et al.* method. We would also like to thank Marc Davis, Avishai Dekel, Alan Heavens, Caleb Scharf, Andy Taylor, Simon White, and David Weinberg for helpful comments and suggestions. We thank the other members of the 1.2 Jy team (M. Davis, J. Huchra, M. Strauss, and A. Yahil) for allowing us to use the redshift data prior to publication. KBF acknowledges a SERC postdoctoral fellowship. OL acknowledges the hospitality of the Hebrew University. YH and SZ acknowledge the hospitality of the Institute of Astronomy.

**TABLE 1**
Mock IRAS CDM Simulation
Peculiar Velocity Comparison[1]
(Distances within 60 $h^{-1}$ Mpc)

| Model | Mean Difference | Mean Absolute Deviation | RMS | Slope |
|---|---|---|---|---|
| *Comparison with True Velocities* | | | | |
| Pure Redshift Space | -73 | 436 | 580 | 1.82 |
| Redshift + Wiener | -52 | 260 | 348 | 1.46 |
| Real[2] Space | -17 | 206 | 283 | 1.18 |
| Real[2] + Wiener | -7 | 147 | 187 | 0.94 |

[1] Velocities in kms$^{-1}$.
[2] Harmonics corrected for redshift distortion.

## A  APPENDIX: ORTHOGONAL RADIAL FUNCTIONS AND BOUNDARY CONDITIONS

Cosmic density fields can be expanded by various orthogonal radial functions such as spherical Bessel functions (Binney & Quinn 1991; Lahav 1994), Laguerre (Lynden-Bell 1991) and Chebyshev polynomials. If our prime motivation were merely to expand the density field, then any of these orthogonal radial functions would be adequate. Yet, we wish to maintain correspondence with the dynamical equations relating the density, velocity, and potential fields. In this case, the spherical Bessel functions become a natural choice since they are not only orthogonal but together with the spherical harmonics form the eigenfunctions of the Laplacian operator which appears in the Poisson equation; as we will see in Appendix C, the Bessel functions lead to particularly simple relationships between the density and potential harmonics.

In order to address the issue of orthogonality more precisely, it is useful to recall the differential equation which defines the spherical Bessel functions (e.g. Jackson 1975, p. 740):

$$\frac{1}{r}\frac{d^2}{dr^2}\left[r j_l(kr)\right] = \left[\frac{l(l+1)}{r^2} - k^2\right] j_l(kr) \quad . \tag{A1}$$

If we multiply both sides of this equation by $j_l(k'r)$ and integrate with respect to $r^2 dr$ from 0 to $R$ we have

$$\int_0^R dr\, r^2 \left\{\frac{1}{r}\frac{d^2}{dr^2}\left[r j_l(kr)\right]\right\} j_l(k'r) = \int_0^R dr\, r^2 \left\{\frac{l(l+1)}{r^2} - k^2\right\} j_l(kr) j_l(k'r) \quad . \tag{A2}$$

Next subtract from this equation the same equation but with $k$ and $k'$ interchanged:

$$\begin{aligned}
(k^2 - k'^2) \int_0^R dr\, r^2 j_l(kr) j_l(k'r) &= \int_0^R dr \left\{ r j_l(k'r) \frac{d^2}{dr^2}\left[r j_l(kr)\right] - r j_l(kr) \frac{d^2}{dr^2}\left[r j_l(k'r)\right] \right\} \\
&= \left[ r j_l(k'r) \frac{d}{dr}\left(r j_l(kr)\right) - r j_l(kr) \frac{d}{dr}\left(r j_l(k'r)\right) \right]\Big|_0^R \\
&= R \left[ (kR) j_l(k'R) j_l'(kR) - (k'R) j_l(kR) j_l'(k'R) \right] \quad .
\end{aligned} \tag{A3}$$

From the last line in Equation A3, we see that the Bessel functions will be orthogonal in the sense that

$$\int_0^R dr\, r^2 j_l(kr) j_l(k'r) = \delta^K_{kk'} C_{ln}^{-1} \quad , \tag{A4}$$

if the radial wavenumbers satisfy the following constraint,

$$A\, j_l'(kR) = B\, \frac{j_l(kR)}{kR} \quad , \tag{A5}$$

where $A$ and $B$ are an arbitrary constants.

From the infinite set of $A$ and $B$s which guarantee orthogonality, we wish to pick those which correspond to a physically well-motivated solution. We assume that the data are given only within a sphere of radius $R$, such that inside the sphere the desired density fluctuation is specified by $\delta(\mathbf{r})$ while for $r > R$ we set $\delta(\mathbf{r}) = 0$. This simply reflects our ignorance about the density field outside the sphere; the fluctuations do not, of course, vanish at large distances. Notice that in the usual case of flux-limited samples the choice of $R$ is somewhat arbitrary and involves an inherent compromise: a small value of $R$ might neglect dynamically important perturbations on scales larger than $R$ while a bigger value of $R$ will increase the statistical noise since the sampling decreases with distance.

Mathematically, the approximation of neglecting fluctuations outside $R$ amounts to solving Poisson's equation for $r < R$ and Laplace's equation for $r \geq R$. The solution to the Poisson equation for $r < R$ can be expressed as a Fourier-Bessel series, while the solution for $r \geq R$ is given by the well known decaying solution to Laplace's equation (e.g. Jackson, p. 90):

$$\psi(\mathbf{r}) = \begin{cases} \sum_{lmn} C_{ln} \psi_{lmn} j_l(k_n r) Y_{lm}(\hat{\mathbf{r}}), & \text{if } r < R \\ \sum_{lm} A_{lm} \left(\frac{R}{r}\right)^{l+1} Y_{lm}(\hat{\mathbf{r}}), & \text{if } r \geq R \end{cases} \quad . \tag{A6}$$

Continuity of $\psi(\mathbf{r})$ at $r = R$ requires that

$$A_{lm} = \sum_n C_{ln}\, \psi_{lmn}\, j_l(k_n R) \quad , \tag{A7}$$

while continuity of the logarithmic derivative, $\frac{d \ln \psi(\mathbf{r})}{d \ln r}$, at $r = R$ further requires that

$$\left.\frac{d \ln j_l(k_n r)}{d \ln r}\right|_{r=R} = -(l+1) \quad . \tag{A8}$$

**TABLE A1**
NORMALIZATION COEFFICIENTS, $C_{ln}$

| Boundary Condition | $k_n$ Constraint | $C_{ln}^{-1}$ |
|---|---|---|
| Density | $j_l(k_n R) = 0$ | $\frac{R^3}{2}[j_l'(k_n R)]^2$ |
| Potential | $j_{l-1}(k_n R) = 0$ | $\frac{R^3}{2}[j_l(k_n R)]^2$ |
| Velocity | $j_l'(k_n R) = 0$ | $\frac{R^3}{2(k_n R)^2}\left((k_n R)^2 - l(l+1)\right)[j_l(k_n R)]^2$ |

From Equation A5 we see that this condition will lead to orthogonality since in this case $A = 1$ and $B = -(l+1)$; the resulting constraint on $k$ is then given by the restriction that $k$ satisfy $j_l'(k_n R) + (l+1)j_l(k_n R)/k_n R = 0$. This relation can be reduced to the condition, $j_{l-1}(k_n R) = 0$ by using the recursion relations for the Bessel functions (e.g. Abramowitz & Stegun, §10.1.19).

Two other possible boundary conditions merit discussion. Perhaps the most natural boundary conditions from a mathematical point of view correspond to Equation A5 with the condition that either $A = 0$ or $B = 0$. The choice $A = 0$ corresponds to setting $\delta(\mathbf{r})$ to zero at $r = R$ which is the same assumption described above; it can be shown, however, that this boundary condition leads to a discontinuous potential (and consequently a discontinuous radial velocity field) at $r = R$. The choice of $B = 0$ corresponds to setting the radial velocity to zero at the boundary. This condition has the undesirable feature of introducing artificial structure ("mirror images") in the density field for $r > R$ in order to compensate for the velocity generated by the matter within $r < R$. Moreover, when the velocity boundary conditions are used in the Fourier-Bessel series (cf. Equation 1), the resulting field will have a vanishing mean value; consequently, if the velocity boundary conditions are used to expand the density field, then one must add the mean density to series expansion given in Equation 1. The values of $C_{ln}$ for the various boundary conditions are given in Table A1.

## B   APPENDIX: RESOLUTION AND THE CHOICE OF $l_{max}$ AND $n_{max}(l)$

In our expansion of the density field we necessarily truncate the summations over the angular and radial modes at some $l_{max}$ and $n_{max}(l)$ respectively. This truncation is an effective smoothing of the density field and limits the amount of small scale structure in the reconstruction; this is analogous to the more commonly employed smoothing of the density by Gaussian or top-hat filters. A convenient way to visualize this smoothing is to consider a single angular and radial mode. The spherical harmonic functions, $Y_{lm}$ have an effective resolution of $\Delta\theta \sim \pi/l$ (cf. Peebles 1980, § 46). The zeros of the Bessel functions, $j_l(z)$, are asymptotically ($z \gg 1$) given by $z_{ln} \simeq \pi(n + l/2)$ where $z_{ln}$ is the $n^{\text{th}}$ zero of $j_l(z)$. Consequently, the resolution of the radial mode is approximately $\Delta r \sim \pi/k_n$ where $k_n = z_{ln}/R$ and $R$ is the radius of the sphere. Thus, a given angular and radial mode roughly divides space into a series of cells with volume $\Delta V \sim (\Delta\theta)^2 r^2 \Delta r = \pi^3/l^2(r^2/k_n)$.

By an appropriate choice for the number of radial modes we can match the effective angular and radial resolution. If the density field is reconstructed from a redshift catalogue of effective depth, $D$, then the $l^{\text{th}}$ order harmonic probes a typical transverse length of $D\Delta\theta$. Equating this with the radial resolution, $\Delta r$, gives the crude condition $k_n \sim l/D$.

Another practical consideration is the total number of harmonic coefficients versus the total number of galaxies used to determine them. Each galaxy has three degrees of freedom corresponding to its position in the sky so a strict upper limit on the number of coefficients used in the expansion of $N$ galaxies is $3N$ (for the *IRAS* 1.2 Jy sample, $N$=5313). The number of angular coefficient for a given $l_{max}$ is $\sum_{l=0}^{l_{max}}(2l+1) = 1 + l_{max}(l_{max} + 2)$. For each angular mode, there will be a set of $n_{max}(l)$ radial modes, where $n_{max}(l)$ is given by the number of zeros in $j_l(x)$ for $x$ less than the maximum cutoff in $k_n R$. There is no simple analytic expression for $n_{max}(l)$ and in general it will depend on the adopted boundary conditions. Table A2 shows the number of expansion coefficients for the potential boundary condition for several values of $l_{max}$ and limiting values of $k_n R$.

Figure 6 gives a illustration of the radial resolution of the Bessel functions for the different boundary conditions for two simple functions. In panels *a*) and *b*) of Figure 6, the function is linear and we show the reconstructed radial distribution for $l = 1$ angular mode. In this case both the potential and velocity distributions closely approximate the function, while the density boundary condition gives a poor representation of the function. In the case of a boxcar distribution (panels *c* and

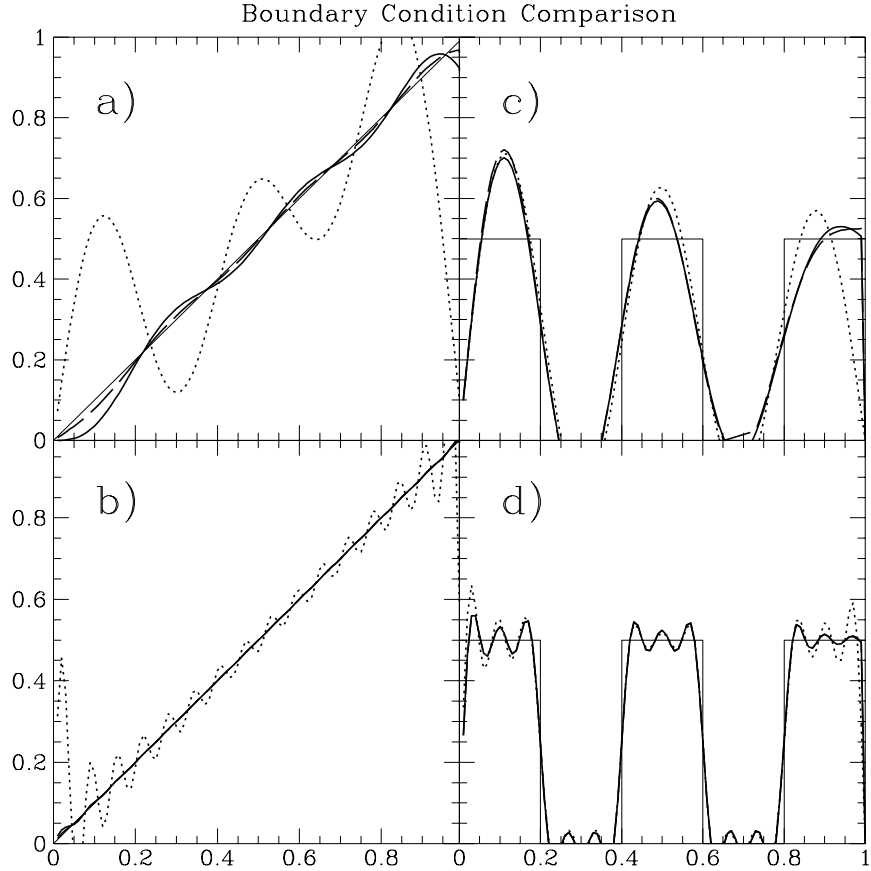

**Figure 6.** Comparison of different boundary conditions for a set of idealized data. a). Reconstruction of the dipole component $l = 1$ when $\rho_{lm}(r) \equiv \sum C_{ln}\rho_{lmn}j_l k_n r = r$ and the radial modes are limited to $k_n R \leq 20$. The light solid curve shows the function to be reconstructed. The reconstructions are performed using the density (dotted line), velocity (dashed), and potential (heavy solid) boundary conditions described in Appendix A. b) Same as in a) with but with higher resolution $k_n R \leq 100$. c). Same as in a) but for a "box car" function. d). Same as in c) but with $k_n R \leq 100$.

**TABLE A2**
Number of Expansion Coefficients
Potential Boundary Conditions

|  | \multicolumn{4}{c}{Maximum of $k_n R$} | | | |
| --- | --- | --- | --- | --- |
| $l_{max}$ | 25 | 50 | 75 | 100 |
| 5 | 226 | 503 | 791 | 1079 |
| 10 | 594 | 1530 | 2483 | 3432 |
| 15 | 912 | 2843 | 4876 | 6905 |

*d*) all three boundary conditions give similar results. The moral is clear: the accuracy of the Fourier-Bessel series for a given choice of boundary conditions will depend on the function being expanded.

There is one drawback to using Bessel functions to expand the density field of a flux-limited galaxy catalogue that arises from their asymptotic properties. For $x \gg l$, the zeros of $j_l(x)$ are equally spaced, while the mean interparticle separation of galaxies in the catalogues steadily increases with distance; consequently the Bessel functions are inefficient representation of the galaxy density field at large distances. One potential way of making the resolution of the Bessel functions adaptive to the galaxy density is to introduce a coordinate transformation which maps the observed galaxy distribution to one with constant interparticle separation. The amounts to introducing a new radial coordinate, $\eta(r)$ which is defined by the condition $\phi(r) d^3 \mathbf{r} = d^3 \eta$ or

$$\phi(r) \, r^2 \, dr \, d\Omega_\mathbf{r} = \eta^2 \, d\eta \, d\Omega_\eta \tag{B1}$$

$$\rightarrow \quad \eta(r) = \left[ 3 \int_0^r dr' \, r'^2 \, \phi(r') \right]^{\frac{1}{3}} \quad . \tag{B2}$$

The integral in Equation B2 is finite as $r \rightarrow \infty$; thus the "$\eta$-space" is finite with equal steps in $\eta$ corresponding to equal increments in the mean interparticle separation. This approach has the advantage of reducing the number of coefficients needed but at the expense of making the formalism for the dynamics much more mathematically cumbersome. In our analysis we have opted for simplicity rather than efficiency and do not apply this transformation.

## C APPENDIX: THE VELOCITY AND POTENTIAL FIELDS IN THE HARMONIC REPRESENTATION

Expressions for the radial velocity field in spherical coordinates in linear theory can be found in Regös & Szalay (1989) and FSL; for completeness we give a brief derivation here. The expressions for the radial velocity are most easily derived by considering the solution to Poisson's equation in spherical coordinates. Poisson's equation is given by

$$\nabla^2 \psi(\mathbf{r}) = 4\pi G \left[ \rho(\mathbf{r}) - \bar{\rho} \right] = \frac{3}{2} \Omega \, \delta(\mathbf{r}) \quad , \tag{C1}$$

where $\delta(\mathbf{r})$ is the fluctuation in the *mass* distribution (we work in units with $H_o$ and the expansion factor set to unity). This equation can be solved by expanding $\psi(\mathbf{r})$ and $\delta(\mathbf{r})$ in Fourier-Bessel series (as in Equation A6) and then recalling that $j_l(k_n r) Y_{lm}(\hat{\mathbf{r}})$ is an eigenfunction of the Laplacian operator, i.e.,

$$\nabla^2 \left[ j_l(k_n r) Y_{lm}(\hat{\mathbf{r}}) \right] = -k_n^2 \, j_l(k_n r) Y_{lm}(\hat{\mathbf{r}}) \quad . \tag{C2}$$

Thus we see from Equation C1 that the harmonics of the potential and density fields are simply related by

$$\psi_{lmn} = -3/2 \Omega \delta_{lmn}/k_n^2 \quad . \tag{C3}$$

In linear theory, the velocity field obeys potential flow with $\mathbf{v}(\mathbf{r}) = -2/3 \Omega^{-0.4} \nabla \psi(\mathbf{r})$. Thus the peculiar velocity field when expressed in terms of the harmonics of the *galaxy* density field, $\delta^R_{lmn} = b \, \delta_{lmn}$ ($b$ being the bias parameter), is given by

$$\mathbf{v}(\mathbf{r}) = \beta \sum_{lmn} C_{ln} \delta^R_{lmn} \frac{1}{k_n^2} \nabla \left[ j_l(k_n r) Y_{lm}(\hat{\mathbf{r}}) \right] \quad , \tag{C4}$$

with $\beta \equiv \Omega^{0.6}/b$.

### C.1 The Radial Velocity Field

The radial component is particularly simple since $\hat{\mathbf{r}} \cdot \nabla = \partial/\partial r$. From Equation C4 the radial velocity field is

$$v_r(\mathbf{r}) = \hat{\mathbf{r}} \cdot \mathbf{v}(\mathbf{r}) \equiv U(\mathbf{r}) = \beta \sum_{lmn} C_{ln} \, \delta^R_{lmn} \frac{j'_l(k_n r)}{k_n} Y_{lm}(\hat{\mathbf{r}}) \quad . \tag{C5}$$

We can use Equation C5 to relate the harmonics of the velocity field, $v_{lmn}$ to those of the density field, $\delta^R_{lmn}$. If we expand the velocity field due to the galaxies with $r < R$ in harmonics then

$$v_r(\mathbf{r}) = \sum_{lmn} C_{ln} \, v_{lmn} \, j_l(k_n r) \, Y_{lm}(\hat{\mathbf{r}}) \quad . \tag{C6}$$

The harmonics are given by the inverse transform which with Equation C5 becomes

$$
\begin{aligned}
v_{lmn} &= \int_{V_R} d^3\mathbf{r}\, Y_{lm}^*(\hat{\mathbf{r}})\, j_l(k_n r)\, v_r(\mathbf{r}) \\
&= \beta \sum_{n'} C_{ln'}\, \delta_{lmn'}^R\, \frac{1}{k_n} \int_0^R dr\, r^2 j_l(k_n r) j_l'(k_{n'} r) \\
&\equiv \beta \sum_{n'} (\Xi_l)_{nn'}\, \delta_{lmn'}^R \quad ,
\end{aligned}
\tag{C7}
$$

where the density to velocity matrix is

$$
(\Xi_l)_{nn'} = \frac{1}{k_{n'}} C_{ln'} \int_0^R dr\, r^2 j_l(k_n r) j_l'(k_{n'} r) \quad . \tag{C8}
$$

Thus, in the transform space, the radial velocity and density fields are related linearly by a matrix multiplication.

Equation C5 is not valid at the origin, so we need to derive an additional expression for $U(\mathbf{0})$. The observer's velocity, $\mathbf{v}(\mathbf{0})$, is given by,

$$
\begin{aligned}
\mathbf{v}(\mathbf{0}) &= \frac{\Omega^{0.6}}{4\pi} \int_{V_R} d^3\mathbf{r}'\, \delta(\mathbf{r}')\, \frac{\mathbf{r}'}{r'^3} \\
&= \frac{\beta}{4\pi} \sum_{lmn} C_{ln}\delta_{lmn}^R \int_0^R dr'\, j_l(k_n r') \int_{4\pi} d\Omega\, \hat{\mathbf{r}}'\, Y_{lm}(\hat{\mathbf{r}}) \\
&= \frac{\beta}{4\pi} \left(\frac{4\pi}{3}\right)^{1/2} \sum_n C_{1n} \left(-\sqrt{2} Re[\delta_{11n}^R]\hat{\mathbf{x}} + \sqrt{2} Im[\delta_{11n}^R]\hat{\mathbf{y}} + Re[\delta_{10n}^R]\hat{\mathbf{z}}\right) (1 - j_0(k_n R)) \quad , \tag{C9}
\end{aligned}
$$

$$\tag{C10}$$

where $Re[a]$ and $Im[a]$ refer to the real and imaginary parts of a complex number, $a$. The last line of Equation C10 gives a convenient Cartesian representation of the observer's velocity, commonly referred to as the "dipole". A useful expression for $U(\mathbf{0})$ can be derived by noting that

$$
\hat{\mathbf{r}} \cdot \int_{4\pi} d\Omega\, \hat{\mathbf{r}}'\, Y_{lm}(\hat{\mathbf{r}}') = \frac{4\pi}{3} \delta_{l1}^K \sum_{m=-1,0,1} Y_{1m}(\hat{\mathbf{r}}) \quad , \tag{C11}
$$

and therefore from the second line in Equation C10

$$
U(\mathbf{0}) = \frac{\beta}{3} \sum_n \sum_{m=-1,0,1} C_{1n}\delta_{1mn}^R (1 - j_0(k_n R)) Y_{1m}(\hat{\mathbf{r}}) \quad . \tag{C12}
$$

### C.2 The Transverse Velocity Field

The expression for the transverse velocity field, $\mathbf{v}_\perp(\mathbf{r})$, due to the matter within $r < R$ can be derived from Equation C4 since

$$
\begin{aligned}
\mathbf{v}_\perp(\mathbf{r}) &= -\mathbf{r} \times [\mathbf{r} \times \mathbf{v}(\mathbf{r})] \\
&= -\beta \sum_{lmn} C_{ln}\, \delta_{lmn}^R\, \frac{1}{k_n^2}\, \mathbf{r} \times [\mathbf{r} \times \nabla (j_l(k_n r) Y_{lm}(\hat{\mathbf{r}}))] \\
&= \beta \sum_{lmn} C_{ln}\, \delta_{lmn}^R\, \frac{j_l(k_n r)}{k_n^2} (i\, \mathbf{r} \times \mathbf{L} Y_{lm}(\hat{\mathbf{r}})) \quad , \tag{C13}
\end{aligned}
$$

where $\mathbf{L}$ is the dimensionless angular momentum operator, $-i\,\mathbf{r} \times \nabla$. The vectors $\mathbf{L} Y_{lm}(\hat{\mathbf{r}})$ are known as spherical vector harmonics (e.g., Morse & Fesbach 1953). They can be derived using methods borrowed from quantum mechanics:

$$
\mathbf{L} Y_{lm}(\hat{\mathbf{r}}) = (\hat{\mathbf{x}} L_x + \hat{\mathbf{y}} L_y + \hat{\mathbf{z}} L_z) Y_{lm}(\hat{\mathbf{r}}) \quad , \tag{C14}
$$

where

$$
\begin{aligned}
L_x &= \frac{1}{2}(L_+ + L_-) \\
L_y &= \frac{-i}{2}(L_+ - L_-) \quad ,
\end{aligned}
\tag{C15}
$$

and

$$\begin{aligned}
L_+ Y_{lm}(\hat{\mathbf{r}}) &= \sqrt{(l-m)(l+m+1)}\, Y_{l,m+1}(\hat{\mathbf{r}}) \\
L_- Y_{lm}(\hat{\mathbf{r}}) &= \sqrt{(l+m)(l-m+1)}\, Y_{l,m-1}(\hat{\mathbf{r}}) \\
L_z Y_{lm}(\hat{\mathbf{r}}) &= m\, Y_{lm}(\hat{\mathbf{r}}) \quad .
\end{aligned} \tag{C16}$$

(cf., Jackson 1975, p. 743).

## D APPENDIX: DERIVATION OF THE RADIAL COUPLING MATRIX

We start by writing an expression for the observable harmonics in redshift space in the *fluid* limit (cf. FSL),

$$\rho^S_{lmn} = \int_{V_S} d^3\mathbf{s}\, \rho_S(\mathbf{s})\, \phi(r)\, j_l(k_n s)\, w(s)\, Y^*_{lm}(\hat{\mathbf{r}}) \quad , \tag{D1}$$

where $V_S$ denotes a spherical integration region in redshift space of radius $R$ and $w(s)$ is an arbitrary weight function used in determining the density. In the application discussed in the text, the weight function was chosen to be $1/\phi(s)$ however in the derivation that follows we leave it as an arbitrary function. The selection function in real space appears in Equation D1 since the probability of a galaxy being in the flux-limited survey depends on its (unknown) distance.

We can convert the above integral in redshift space to real space using linear theory. In the absence of orbits undergoing shell crossing, there is a unique mapping from real to redshift space which conserves the number of galaxies:

$$\rho_S(\mathbf{s})\, d^3\mathbf{s} = \rho_R(\mathbf{r})\, d^3\mathbf{r} \quad . \tag{D2}$$

In the linear theory limit, we expect Equation D2 to hold; we proceed by applying this mapping to Equation D1,

$$\rho^S_{lmn} = \int_{V_S} d^3\mathbf{r}\, \rho_R(\mathbf{r})\phi(r) j_l(k_n s) w(s) Y^*_{lm}(\hat{\mathbf{r}}) \quad . \tag{D3}$$

We can use the redshift-distance relation to expand any function of redshift in a Taylor series that depends on distance. From the relation

$$s = r + U(\mathbf{r}) - U(\mathbf{0}) \quad , \tag{D4}$$

we have for an arbitrary function of redshift,

$$f(s) = f(r) + \frac{df(r)}{dr}\left(U(\mathbf{r}) - U(\mathbf{0})\right) + \mathcal{O}\left[(U(\mathbf{r}) - U(\mathbf{0}))^2\right] \quad . \tag{D5}$$

We note that by including the term $U(\mathbf{0})$ in Equation D4 we are assuming that the redshifts are in the LG frame; if one works with redshifts in the CMB frame, then this term should be dropped.

The expansion of the function $j_l(k_n s) w(s)$ in Equation D3 using Equation D5 yields

$$\begin{aligned}
\rho^S_{lmn} &= \int_{V_S} d^3\mathbf{r}\, \rho_R(\mathbf{r})\phi(r) j_l(k_n s) w(s) Y^*_{lm}(\hat{\mathbf{r}}) = \bar{\rho} \int_{V_S} d^3\mathbf{r}\, \left(1 + \delta^R(\mathbf{r})\right) \phi(r)\, j_l(k_n s) w(s) Y^*_{lm}(\hat{\mathbf{r}}) \\
&= \bar{\rho} \int_{V_R} d^3\mathbf{r}\, \phi(r) Y^*_{lm}(\hat{\mathbf{r}})\left[\left(1 + \delta^R(\mathbf{r})\right) j_l(k_n r)\, w(r) + \left(\frac{d}{dr}\left(j_l(k_n r)\, w(r)\right)\right)(U(\mathbf{r}) - U(\mathbf{0}))\right] + \\
&\quad \int_{V_S - V_R} d^3\mathbf{r}\, \phi(r) w(r) j_l(k_n r) Y^*_{lm}(\hat{\mathbf{r}}) \quad ,
\end{aligned} \tag{D6}$$

where we have retained terms to first order in both $\delta(\mathbf{r})$ and $U(\mathbf{r})$. The last term in the above equation is a surface term that arises because the spherical integration region in redshift space is deformed in real space.

From Equation D6, we see that the first order expression for $\rho^S_{lmn}$ can be broken down into three distinct terms:

$$\rho^S_{lmn} = \bar{\rho}\left(\delta^R_{lmn} + \mathbf{M}_{lmn} + \mathbf{D}_{lmn}\right) \quad , \tag{D7}$$

where

$$\delta^R_{lmn} \equiv \int_{V_R} d^3\mathbf{r}\, \delta^R(\mathbf{r})\, \phi(r)\, w(r)\, j_l(k_n r)\, Y^*_{lm}(\mathbf{r}) \qquad \underline{\text{real space contribution}}$$

$$\mathbf{M}_{lmn} \equiv \int_{V_R} d^3\mathbf{r}\, \phi(r)\, w(r)\, j_l(k_n r)\, Y^*_{lm}(\hat{\mathbf{r}}) \qquad \underline{\text{monopole term}}$$

$$\mathbf{D}_{lmn} \equiv \int_{V_R} d^3\mathbf{r}\, \phi(r)\, Y_{lm}^*(\hat{\mathbf{r}})\, \left[\frac{d}{dr}\left(w(r)\, j_l(k_n r)\right)\right] (U(\mathbf{r}) - U(\mathbf{0}))$$
$$+ \int_{V_S - V_R} d^3\mathbf{r}\, \phi(r)\, w(r)\, j_l(k_n r)\, Y_{lm}^*(\hat{\mathbf{r}}) \qquad \underline{\text{distortion term}} \quad .$$
(D8)

Let's examine the distortion term first. We begin by integrating the first integral in $\mathbf{D}_{lmn}$ by parts. This yields

$$\int_{4\pi} d\Omega\, Y_{lm}^*(\hat{\mathbf{r}}) \int_0^R dr \left\{ \frac{d}{dr}\left[r^2 \phi(r) w(r) j_l(k_n r) (U(\mathbf{r}) - U(\mathbf{0}))\right] - w(r) j_l(k_n r) \frac{d}{dr}\left[r^2 \phi(r) (U(\mathbf{r}) - U(\mathbf{0}))\right] \right\}$$

$$= \int_{4\pi} d\Omega\, Y_{lm}^*(\hat{\mathbf{r}})\, R^2\, \phi(R)\, w(R)\, j_l(k_n R)\, (U(\mathbf{R}) - U(\mathbf{0}))$$

$$- \int_{V_R} d^3\mathbf{r}\, Y_{lm}^*(\hat{\mathbf{r}})\, \phi(r)\, w(r)\, j_l(k_n r)\, \left[\left(2 + \frac{d\ln\phi(r)}{d\ln r}\right)\left(\frac{U(\mathbf{r}) - U(\mathbf{0})}{r}\right) + \frac{dU(\mathbf{r})}{dr}\right] \quad .$$
(D9)

Next, consider the second integral in $\mathbf{D}_{lmn}$:

$$\int_{V_S - V_R} d^3\mathbf{r}\, \phi(r) w(r) j_l(k_n r) Y_{lm}^*(\hat{\mathbf{r}}) = \int_{4\pi} d\Omega\, Y_{lm}^*(\hat{\mathbf{r}}) \int_R^{R - U(\mathbf{R}) + U(\mathbf{0})} dr\, r^2\, \phi(r)\, w(r)\, j_l(k_n r)$$

$$\approx -\int_{4\pi} d\Omega\, Y_{lm}^*(\hat{\mathbf{r}}) R^2 \phi(R) w(R) j_l(k_n R)\, (U(\mathbf{R}) - U(\mathbf{0})) \quad .$$
(D10)

Comparing Equations D9 and D10 we see that there is a fortuitous cancellation of the surface terms. The remaining terms give a distortion identical to that in Kaiser (1987) only spherically transformed,

$$\mathbf{D}_{lmn} = -\int_{V_R} d^3\mathbf{r}\, Y_{lm}^*(\hat{\mathbf{r}})\, \phi(r)\, w(r)\, j_l(k_n r)\, \left[\frac{dU(\mathbf{r})}{dr} + \left(\frac{U(\mathbf{r}) - U(\mathbf{0})}{r}\right)\left(2 + \frac{d\ln\phi(r)}{d\ln r}\right)\right] \quad .$$
(D11)

In Appendix C (see also FSL; Regös & Szalay 1989) we show that the radial velocity field (in linear theory) in spherical coordinates is given by

$$U(\mathbf{r}) = \beta \sum_{lmn} C_{ln} \delta_{lmn}^R \frac{j_l'(k_n r)}{k_n} Y_{lm}(\mathbf{r})$$

$$U(\mathbf{0}) = \frac{\beta}{3} \sum_{lmn} \delta_{l1}^K C_{ln} \delta_{lmn}^R \int_0^R dr\, j_l(k_n r) Y_{lm}(\mathbf{r}) \quad .$$
(D12)

Substituting these relations into Equation D11 and performing the angular integration over $d\Omega$ yields:

$$\mathbf{D}_{lmn} = -\beta \sum_{n'} C_{ln'} \delta_{lmn'}^R \int_0^R dr\, r^2 \phi(r) w(r) j_l(k_n r) \left[j_l''(k_{n'} r) + \left(\frac{j_l'(k_{n'} r)}{k_{n'} r} - \frac{\delta_{l1}^K}{3} \frac{1}{r} \int_0^R dx\, j_l(k_n' x)\right)\left(2 + \frac{d\ln\phi(r)}{d\ln r}\right)\right] \quad .$$
(D13)

In our reconstruction, we took $w(r) = 1/\phi(r)$. In this case the monopole contribution is given by

$$\mathbf{M}_{lmn} = \int_{V_R} d^3\mathbf{r}\, j_l(k_n r)\, Y_{lm}^*(\hat{\mathbf{r}}) = \sqrt{4\pi} \int_{V_R} d^3\mathbf{r}\, j_l(k_n r)\, Y_{lm}^*(\hat{\mathbf{r}}) Y_{00}(\hat{\mathbf{r}})$$

$$= \sqrt{4\pi} \int_0^R dr\, r^2 j_0(k_n r) = \left(\sqrt{4\pi} R^3\right) \left(\frac{j_1(k_n R)}{k_n R}\right) \delta_{l0}^K \delta_{m0}^K \quad ,$$
(D14)

which is the result quoted in Equation 5. The redshift harmonics of the fluctuation field, $\delta_{lmn}^S$, are thus given by,

$$\delta_{lmn}^S = \frac{\rho_{lmn}^S}{\bar{\rho}} - \mathbf{M}_{lmn} \quad .$$
(D15)

This result can be combined with Equation D13 for the distortion term to express Equation D7 in a way which clearly shows the coupling of the radial modes in redshift space. Let

$$\delta_{lmn}^S = \sum_{n'} (\mathbf{Z}_l)_{nn'} \delta_{lmn'}^R \quad , \tag{D16}$$

where we have defined the coupling matrix, for a given $l$ and $\phi(r)w(r) = 1$ as

$$(\mathbf{Z}_l)_{nn'} = \delta_{nn'}^K - \beta \int_0^R dr\, r^2 j_l(k_n r) \left[ j_l''(k_{n'} r) + \left( \frac{j_l'(k_{n'} r)}{k_{n'} r} - \frac{\delta_{l1}^K}{3} \frac{1}{r} \int_0^R dx\, j_l(k_n' x) \right) \left( 2 + \frac{d \ln \phi(r)}{d \ln r} \right) \right] \quad . \tag{D17}$$

## E  APPENDIX: MODEL SIGNAL AND NOISE MATRICES

In evaluating the WF, we need a model for the expected variance in the signal and noise. For the particular case we are considering, this amounts to computing the expected signal and "shot noise" of the harmonics given by

$$\hat{\delta}_{lmn}^R = \bar{\rho} \int_{V_R} d^3\mathbf{r}\, \phi(r)\, w(r)\, \delta^R(\mathbf{r})\, j_l(k_n r)\, Y_{lm}^*(\hat{\mathbf{r}}) \quad . \tag{E1}$$

Unlike previous Appendices, we are treating the fluctuation field in Equation E1 as that due to a discrete set of galaxies. Notice that we must compute the expectation value of the estimator in *real* space since it is the true underlying signal and noise in real space that appears the WF formulation.

The matrix which appears in the WF is $\langle \hat{\delta}_{lmn}^R \hat{\delta}_{l'm'n'}^{R\dagger} \rangle$. This expectation value will contain both the signal and the noise arising from the discrete nature of the galaxies. Since we assume full sky coverage, the expectation value is diagonal in $(l,m)$; consequently the matrix for a given value of $l$ is a square matrix with dimensions $n_{max}(l) \times n_{max}(l)$, where $n_{max}(l)$ is the number of radial modes in the expansion for the given $l$.

The expectation value can be computed by recalling that for a point distribution the expectation value of the density fluctuation is given by

$$\langle \delta^R(\mathbf{r}_1) \delta^R(\mathbf{r}_2) \rangle^d = \xi(|\mathbf{r}_1 - \mathbf{r}_2|) + \frac{1}{\bar{\rho}\phi(r)} \delta_D^{(3)}(\mathbf{r}_1 - \mathbf{r}_2) \quad , \tag{E2}$$

where the superscript $d$ denotes the ensemble average of a discrete point set and the subscript $D$ denote the Dirac delta function (cf., Bertschinger 1992). The term in the above expression involving the correlation function, $\xi(r)$, leads to the expression for the expected signal, while the term involving the delta function represents the shot noise. Therefore if we square Equation E1 and take its expectation value using Equation E2 we arrive at the following expressions for the signal and noise matrices (cf., § 4):

$$(\mathbf{S}_l)_{nn'} = \bar{\rho}^2 \int_{V_R} d^3\mathbf{r}_1\, d^3\mathbf{r}_2\, \phi(r_1) \phi(r_2) w(r_1) w(r_2) j_l(k_n r_1) j_l(k_{n'} r_2) Y_{lm}^*(\hat{\mathbf{r}}_1) Y_{lm}(\hat{\mathbf{r}}_2) \xi(|\mathbf{r}_1 - \mathbf{r}_2|) \quad , \tag{E3}$$

and

$$(\mathbf{N}_l) = \bar{\rho} \int_{V_R} dr\, r^2\, \phi(r) [w(r)]^2 j_l(k_n r) j_l(k_{n'} r) \quad . \tag{E4}$$

We can simplify Equation E3 by writing the two-point correlation function in terms of the power spectrum, $P(k)$. In spherical coordinates the correlation function is related to the power spectrum by

$$\begin{aligned} \xi(|\mathbf{r}_1 - \mathbf{r}_2|) &= \frac{1}{(2\pi)^3} \int d^3\mathbf{k}\, P(k)\, e^{-i\mathbf{k}\cdot(\mathbf{r}_1 - \mathbf{r}_2)} \\ &= \frac{2}{\pi} \sum_{lm} \sum_{l'm'} Y_{lm}(\hat{\mathbf{r}}_1) Y_{l'm'}^*(\hat{\mathbf{r}}_2) \int d^3\mathbf{k}\, P(k)\, j_l(kr_1) j_{l'}(kr_2) Y_{lm}^*(\hat{\mathbf{k}}) Y_{l'm'}(\hat{\mathbf{k}}) \\ &= \frac{2}{\pi} \sum_{lm} Y_{lm}(\hat{\mathbf{r}}_1) Y_{lm}^*(\hat{\mathbf{r}}_2) \int_0^\infty dk\, k^2 P(k) j_l(kr_1) j_l(kr_2) \quad , \end{aligned} \tag{E5}$$

where the second line follows from the Rayleigh expansion of the exponential in spherical waves,

$$e^{i\mathbf{k}\cdot\mathbf{r}} = 4\pi \sum i^l j_l(kr) Y_{lm}^*(\hat{\mathbf{r}}) Y_{lm}(\hat{\mathbf{k}}) \quad . \tag{E6}$$

Substitution of Equation E5 into Equation E3 yields,

$$(\mathbf{S}_l)_{nn'} = \frac{2}{\pi} \bar{\rho}^2 \int_0^\infty dk\, k^2 P(k) \int_0^R dr_1\, r_1^2 \phi(r_1) w(r_1) j_l(k_n r_1) j_l(kr_1) \int_0^R dr_2\, r_2^2 \phi(r_2) w(r_2) j_l(k_{n'} r_2) j_l(kr_2) \quad . \tag{E7}$$

The multiple integral in Equation E7 can be simplified if we restrict the analysis to well sampled modes, i.e., to waves which satisfy $k_n R \gg 1$. In this limit, the spherical Bessel functions will oscillate rapidly and the integration over $k$ will be sharply peaked about $r_1 = r_2$. Therefore, a good approximation is to factor $\phi(r_2) w(r_2)$ out of the integral over $r_2$ and put it in the integral over $r_1$; this same approximation appears in the small angle approximation relation between the angular and real space correlation function (Limber 1954) as well as in the derivation of the weight function which yields a minimum variance estimate of the mean density (Davis & Huchra 1982). Of course, the approximation is exact in the special case $w(r)\phi(r) = 1$. With this approximation, Equation E7 becomes

$$
\begin{aligned}
(\mathbf{S}_l)_{nn'} &\simeq \frac{2}{\pi} \bar{\rho}^2 \int_0^\infty dk\, k^2 P(k) \int_0^R dr_1\, r_1^2\, \phi^2(r_1) w^2(r_1)\, j_l(k_n r_1)\, j_l(kr_1) \int_0^R dr_2\, r_2^2\, j_l(k_{n'} r_2)\, j_l(kr_2) \\
&\simeq \frac{2}{\pi} \bar{\rho}^2 \int_0^\infty dk\, k^2 P(k) \int_0^R dr_1\, r_1^2\, \phi^2(r_1) w^2(r_1)\, j_l(k_n r_1)\, j_l(kr_1) \int_0^\infty dr_2\, r_2^2\, j_l(k_{n'} r_2)\, j_l(kr_2) \\
&= \frac{2}{\pi} \bar{\rho}^2 \int_0^\infty dk\, k^2 P(k) \int_0^R dr_1\, r_1^2\, \phi^2(r_1) w^2(r_1) j_l(k_n r_1) j_l(kr_1) \left\{ \frac{\pi}{2} \frac{1}{k^2} \delta^{(1)}(k - k_{n'}) \right\} \\
&= \bar{\rho}^2\, P(k_{n'}) \int_0^R dr_1\, r_1^2\, \phi^2(r_1) w^2(r_1)\, j_l(k_n r_1)\, j_l(k_{n'} r_1) \\
&= \bar{\rho}^2\, P(k_n)\, C_{ln}^{-1}\, \delta^K_{nn'} \qquad \text{if } w(r) = 1/\phi(r) \quad . 
\end{aligned}
\tag{E8}
$$

## F APPENDIX: SCATTER IN THE RECONSTRUCTED FIELDS

The WF formalism gives an estimate of the expected scatter in the reconstructed fields. First, consider the reconstructed density field. We have an estimate of the harmonics of the density field, $\hat{\delta}^R_{lmn}$ which are related to the underlying true harmonics, $\delta^R_{lmn}$, by

$$
\begin{aligned}
\hat{\delta}^R_{lmn} &= \mathbf{S}_l\, [\mathbf{S}_l + \mathbf{N}_l]^{-1}\, \mathbf{Z}_l^{-1}\, \delta^S_{lmn} \\
&= \mathbf{S}_l\, [\mathbf{S}_l + \mathbf{N}_l]^{-1}\, \mathbf{Z}_l^{-1}\, \mathbf{Z}_l \left( \delta^R_{lmn} + \epsilon_{lmn} \right) \\
&\equiv \mathbf{F}_l \left( \delta^R_{lmn} + \epsilon_{lmn} \right) \quad ,
\end{aligned}
\tag{F1}
$$

where $\epsilon_{lmn}$ represents the shot noise contribution and $\mathbf{F}_l = \mathbf{S}_l(\mathbf{S}_l + \mathbf{N}_l)^{-1}$. The scatter in the density field is given by

$$
\begin{aligned}
\langle \Delta\delta(\mathbf{r}) \rangle^2 &= \langle \left[ \hat{\delta}(\mathbf{r}) - \delta(\mathbf{r}) \right]^2 \rangle \\
&= \sum_{lmn} \sum_{l'm'n'} \langle \left( \hat{\delta}^R_{lmn} - \delta^R_{lmn} \right) \left( \hat{\delta}^R_{l'm'n'} - \delta^R_{l'm'n'} \right)^\dagger \rangle C_{ln} C_{l'n'}\, j_l(k_n r)\, j_{l'}(k_{n'} r)\, Y^*_{lm}(\hat{\mathbf{r}})\, Y_{l'm'}(\hat{\mathbf{r}}) \\
&= \frac{1}{4\pi} \sum_{lnn'} (2l+1) \langle \left( \hat{\delta}^R_{lmn} - \delta^R_{lmn} \right) \left( \hat{\delta}^R_{lmn'} - \delta^R_{lmn'} \right)^\dagger \rangle C_{ln} C_{ln'}\, j_l(k_n r)\, j_l(k_{n'} r) \quad .
\end{aligned}
\tag{F2}
$$

The third line of Equation F2 follows from the addition theorem of the angular harmonics. If Equation F1 is substituted into Equation F2 with the assumption that the signal and noise are uncorrelated, i.e. $\langle \delta^R_{lmn} \epsilon^\dagger_{lmn} \rangle = 0$, then one can derive a compact expression for the scatter,

$$
\langle \Delta\delta(\mathbf{r}) \rangle^2 = \frac{1}{4\pi} \sum_{lnn'} (2l+1)\, [(\mathbf{I} - \mathbf{F}_l)\, \mathbf{S}_l]_{nn'}\, C_{ln} C_{ln'}\, j_l(k_n r)\, j_l(k_{n'} r) \quad ,
\tag{F3}
$$

where $\mathbf{I}$ is the identity matrix. Notice in particular if the noise vanishes, then $\mathbf{F}_l = \mathbf{I}$ and the scatter in the reconstructed density field goes to zero.

The scatter in the reconstructed radial velocity field can be computed in an analogous way to Equation F3 by recalling Equation 20 for the radial velocity field. The result is given by

$$
\begin{aligned}
\langle \Delta v_r(\mathbf{r}) \rangle^2 &= \langle \left[ \hat{v}_r(\mathbf{r}) - v_r(\mathbf{r}) \right]^2 \rangle \\
&= \frac{\beta^2}{4\pi} \sum_{lnn'} (2l+1)\, [(\mathbf{I} - \mathbf{F}_l)\, \mathbf{S}_l]_{nn'}\, C_{ln} C_{ln'}\, \frac{j'_l(k_n r)\, j'_l(k_{n'} r)}{k_n k_{n'}} \quad .
\end{aligned}
\tag{F4}
$$